\documentclass[journal]{IEEEtran}

\usepackage[labelformat=simple]{subcaption}

\usepackage{amsmath,amsfonts}
\usepackage{algorithmic}
\usepackage{algorithm}
\usepackage{array}
\usepackage{textcomp}
\usepackage{graphicx}
\usepackage{caption}
\usepackage{subcaption}
\usepackage{stfloats}
\usepackage{url}
\usepackage{verbatim}
\usepackage{cite}
\usepackage{threeparttable}
\usepackage{float}
\usepackage{booktabs}
\usepackage{multirow}
\usepackage{lipsum,mwe,cuted}
\usepackage{cuted}
\usepackage{afterpage}
\usepackage{adjustbox}
\usepackage{soul}
\usepackage[citecolor=blue, colorlinks]{hyperref}
\usepackage{xcolor}
\newcommand{\cmmnt}[1]{}

\newcommand{\HXADD}[1]{\textcolor{black}{#1}}
\newcommand{\HXDEL}[1]{} 

\begin{document}

\title{Rendering-Oriented 3D Point Cloud Attribute Compression using Sparse Tensor-based Transformer}

\author{Xiao Huo, Junhui Hou,~\IEEEmembership{Senior Member,~IEEE,}, Shuai Wan,~\IEEEmembership{Member,~IEEE,} and Fuzheng Yang,~\IEEEmembership{Member,~IEEE}

\thanks{Xiao Huo and Fuzheng Yang are with the School of Telecommunication Engineering, Xidian University, Xi’an 710071, China (e-mail: xhuo@stu.xidian.edu.cn
; fzhyang@mail.xidian.edu.cn).}
\thanks{Junhui Hou is with the Department of Computer Science, City University of Hong Kong, Hong Kong SAR (e-mail: jh.hou@cityu.edu.hk).}
\thanks{Shuai Wan is with the School of Electronics and Information, Northwestern Polytechnical University, Xi’an 710129, China, and also with the School of Engineering, Royal Melbourne Institute of Technology, Melbourne, VIC 3001, Australia (e-mail: swan@nwpu.edu.cn).}

}



\maketitle

\begin{abstract}
The evolution of 3D visualization techniques has fundamentally transformed how we interact with digital content. At the forefront of this change is point cloud technology, offering an immersive experience that surpasses traditional 2D representations. However, the massive data size of point clouds presents significant challenges in data compression. Current methods for lossy point cloud attribute compression (PCAC) generally focus on reconstructing the original point clouds with minimal error. However, for point cloud visualization scenarios, the reconstructed point clouds with distortion still need to undergo a complex rendering process, which affects the final user-perceived quality. In this paper, we propose an end-to-end deep learning framework that seamlessly integrates PCAC with differentiable rendering, denoted as rendering-oriented PCAC (RO-PCAC), directly targeting the quality of rendered multiview images for viewing. In a differentiable manner, the impact of the rendering process on the reconstructed point clouds is taken into account. Moreover, we characterize point clouds as sparse tensors and propose a sparse tensor-based transformer, called SP-Trans. By aligning with the local density of the point cloud and utilizing an enhanced local attention mechanism, SP-Trans captures the intricate relationships within the point cloud, further improving feature analysis and synthesis within the framework. Extensive experiments demonstrate that the proposed RO-PCAC achieves state-of-the-art compression performance, compared to existing reconstruction-oriented methods, including traditional, learning-based, and hybrid methods. The code will be released at https://github.com/net-F/RO-PCAC.git.

\end{abstract}


\section{Introduction}
\IEEEPARstart{N}{owdays}, the advent of 3D visualization techniques is revolutionizing the way we interact with digital content, offering a more immersive experience compared to traditional 2D formats. This technological shift has unlocked new possibilities for multimedia applications, including immersive communication~\cite{7434610}, virtual and augmented reality experiences~\cite{10.1145/3145690.3145729}, and the preservation of cultural artifacts~\cite{7123038}. A key representation in 3D visualization is the point cloud, which consists of points in 3D space, often accompanied by attributes such as color and surface properties. The fidelity of a visual scene is closely tied to the number of points in the point cloud, which can range from thousands to billions, posing significant challenges for data management. To address this, various compression techniques~\cite{9503405,9500191,10024999,9957096,8816692,8949735} have been developed to reduce point cloud data size while maintaining quality. In line with the push of the industry for efficient data handling, standardization bodies like the Joint Photographic Experts Group (JPEG) and the Moving Picture Experts Group (MPEG) have recognized the importance of point cloud data formats and have embarked on creating standards for their compression. This has led to the development of the Geometry-based Point Cloud Compression (G-PCC) and Video-based Point Cloud Compression (V-PCC) standards~\cite{8571288,8700674,8945224}, both designed to meet the specific compression requirements of point cloud data. In addition, numerous deep learning-based methods~\cite{9418789,wang2022sparse,huang2020octsqueeze,que2021voxelcontext,fu2022octattention,song2023efficient,deeppcac,9191180,sparseW,10313579,10474112} have been developed for point cloud compression, showing promising performance compared to these traditional approaches.

In point cloud visualization scenarios, the content rendered on the screen represents the final output seen by the end-user. 
Considering the compression process, the transformation of raw point clouds into visually coherent images is a complex process. Critical factors include the accuracy of point positioning, the richness of attribute information, and the effectiveness of rendering algorithms. Javaheri et al. investigated the impact of rendering on multiple MPEG point cloud coding solutions, motivated by the consideration that the perceived quality of point cloud data highly depends on the rendering solution~\cite{9257015}.  Despite this, current point cloud compression methods focus on preserving the accuracy of point locations and associated attributes after reconstruction, often overlooking the effect of the rendering process. 

In this paper, we propose to integrate the compression and rendering modules within a single deep-learning framework to achieve an optimal trade-off between compression efficiency and the quality of rendered multiview images. The key feature of our framework, namely RO-PCAC, is the differentiability of the rendering module. After the rate-distortion loss is calculated, the rendering module generates necessary gradients that guide the network to jointly minimize bitrate and optimize image quality. The rate-distortion loss includes the estimated bitrate and the image error between the original and rendered multiview images. Additionally, inspired by the success of point cloud transformers~\cite{Zhao_2021_ICCV,NEURIPS2022_d78ece66,guo2021pct} and sparse tensor-based convolution~\cite{choy20194d,gwak2020gsdn,MLSYS2022_c48e8203,Tang_2023_CVPR}, we propose a sparse tensor-based transformer, termed SP-Trans, to enhance feature analysis and synthesis. It adaptively constructs local neighborhoods within a fixed-size 3D window, aligning with local point density and ensuring a consistent number of neighbors for the enhanced local self-attention mechanism. Cosine similarity is employed within the local self-attention layer to handle varying point numbers and ensure its functionality in sparse regions. This design reduces computational complexity while leveraging the ability of the transformer to capture complex relationships and long-range dependencies between points. RO-PCAC outperforms state-of-the-art compression methods, including the
traditional, learning-based, and hybrid approaches, on benchmark datasets such as 8i Voxelized Full Bodies (8iVFB)~\cite{8ivfb} and Owlii dynamic human mesh (Owlii)~\cite{owlii}.

In summary, the main contributions of this paper are:
\begin{enumerate}

\item we make the first attempt to construct a generic rendering-oriented compression framework for the color attribute of 3D point cloud data, termed RO-PCAC. RO-PCAC consists of a learning-based point cloud attribute compression module and a differentiable rendering module, enabling end-to-end training to jointly minimize the bitrate and optimize the quality of rendered multiview images. Furthermore, RO-PCAC supports the modular design for both the compression and rendering components.

\item  we propose a sparse tensor-based transformer,  SP-Trans, which captures the attribute relationships within local regions of point clouds and enhances the feature representation. SP-Trans adaptively constructs local neighborhoods and leverages a local self-attention mechanism, supported by cosine similarity, to process the density-varied regions, thereby improving compression efficiency.
    
\end{enumerate}
The remainder of this paper is organized as follows: Section \ref{sec:RW} provides a review of related works. Section \ref{sec:proposed method} describes the motivation of the proposed RO-PCAC and details the framework. Section \ref{sec:exp} first presents the training and testing setup, followed by objective and subjective quality comparison results, along with ablation studies. Finally, concluding remarks are offered in Section \ref{sec:con}.

\section{Related Work}
\label{sec:RW}
In this section, we primarily review point cloud geometry and attribute compression, covering both traditional, learning-based and hybrid methods. In addition, we discuss point cloud-based differentiable rendering techniques.

\subsection{Point Cloud Geometry Compression}
Traditional point cloud geometry coding has mature frameworks~\cite{9503405,9500191,9957096,8816692,8949735,8571288,8700674,8945224}, mainly dominated by MPEG G-PCC and V-PCC standards~\cite{8571288,8700674,8945224}. These frameworks have set the benchmark for efficient compression techniques. The MPEG G-PCC codec, for instance, employs an octree-based partitioning strategy to hierarchically structure point clouds~\cite{schnabel2006octree}, which allows for efficient compression by encoding the octree's occupancy. G-PCC also incorporates a surface modeling technique~\cite{8571288} for point clouds with dense sampling. Surface approximation is achieved through the use of triangular meshes, or 'trisoup', which is handled by a dedicated geometry encoding scheme. In a divergent methodology, V-PCC begins the compression process by converting the data from three-dimensional to two-dimensional space, enabling the use of established video encoding standards such as High-Efficiency Video Coding (HEVC)~\cite{sze2014high} for encoding planes and depth information across a sequence. Currently, G-PCC currently stands as a leading protocol in terms of compression efficiency for static point cloud data.

The evolution of deep learning has advanced the development of novel compression methodologies tailored for point cloud geometry. These methodologies often draw inspiration from the architectural blueprints of 2D image compression models~\cite{balle2016end,balle2018variational,minnenbt18}. Notably, Wang et al. have demonstrated exceptional compression performance, due to the sparse convolution~\cite{Choy_2019_CVPR} based backbone, which accommodates the sparse characteristics of point cloud data~\cite{9418789,wang2022sparse}. Conversely, studies by Huang et al., Que et al., Fu et al. and Song et al. have pursued octree-based approaches to point cloud compression~\cite{huang2020octsqueeze,que2021voxelcontext,fu2022octattention,song2023efficient}. Huang et al. introduced a hierarchical multi-layer perception (MLP)-based framework to predict the probabilities of the occupancy generated by octree construction~\cite{huang2020octsqueeze}. Then these probabilities are fed to an arithmetic coder to compress the occupancy. Que et al. proposed VoxelContext-Net, a deep learning framework that integrates octree and voxel-based methods for point cloud compression~\cite{que2021voxelcontext}. The framework uses a local voxel representation of each node to extract spatial context for probability prediction and occupancy compression. Fu et al. proposed OctAttention, which leverages large-scale contexts for point cloud compression~\cite{fu2022octattention}. The context incorporates sibling and ancestor nodes' information to enhance the predictability of each node's occupancy, and employs an attention mechanism to focus on correlated nodes within the context. Song et al. presented an efficient hierarchical entropy model (EHEM), which incorporates a hierarchical attention that maintains a global receptive field while operating with linear complexity relative to the context scale~\cite{song2023efficient}. Additionally, a grouped context is introduced to facilitate parallel decoding, thus mitigating the decoding latency issues associated with autoregressive models. Zhang et al. parameterized unstructured point sets into regular images dubbed as deep geometry image (DeepGI)~\cite{zhang2022reggeonet}, such that spatial coordinates of unordered points are recorded in three-channel grid pixels. By applying a mature image codec on DeepGI, efficient point cloud geometry compression can be achieved.

\subsection{Traditional Point Cloud Attribute Compression}
Traditional point cloud attribute compression techniques, rooted in signal processing algorithms, provide solid theoretical foundations and ensure high reliability in data compression.

Zhang et al. introduced a pioneering graph transform-based method (GFT) for point cloud attribute compression. This approach leverages the spectral properties of graphs to achieve compression~\cite{7025414}. Its strength lies in capturing the spatial correlations by constructing sub-graphs and associating them with Laplacian matrices to effectively encode the color information of point clouds. Yet, GFT may struggle with point clouds that have complex or irregular topologies. Its practicality is hindered by the computationally intensive task of repeatedly solving eigen-decompositions, which makes it unsuitable for real-time applications. 
Queiroz and Chou proposed a Region-Adaptive Hierarchical Transform (RAHT) with employing a hierarchical sub-band transform~\cite{7482691}. It is adaptive and akin to a modified Haar wavelet, coupled with arithmetic coding that assumes Laplace distributions for the coefficients of each sub-band. This method is advantageous due to its computational efficiency, which enables real-time processing, and its comparable performance to state-of-the-art methods in terms of rate-distortion. 
RAHT has been adopted into the MPEG G-PCC standardization~\cite{g_pcc} as one of the core transformation parts for attribute compression. The standardization process has focused on enhancing efficiency and applicability. Improvements in entropy coding, as seen in G-PCC version 6 (TMC13v6), have optimized the bit-rate allocation for RAHT coefficients, leading to better compression with less loss of quality. The latest enhancements, including the prediction of RAHT coefficients in TMC13v23, have further improved the compression efficiency. This predictive approach leverages the spatial and statistical correlations within point clouds to reduce the amount of information that needs to be encoded, achieving state-of-the-art performance in point cloud attribute compression.

\subsection{Learning-based Point Cloud Attribute Compression}
Recent advancements in deep learning have introduced neural networks capable of discerning complex data patterns and retaining essential information from data, potentially outperforming traditional methods in compression efficiency.

Sheng et al. pioneered an end-to-end compression pipeline for point cloud attributes, termed Deep-PCAC~\cite{deeppcac}. Deep-PCAC utilizes a variational autoencoder that directly encodes and decodes point cloud attributes. A second-order point-based convolution is introduced to widen the receptive field of points and augment the feature extraction. This approach achieves competitive performance compared to traditional methods like RAHT-RLGR~\cite{9191183} but falls short when compared to the state-of-the-art MPEG G-PCC reference software TMC13~\cite{g_pcc}.
Quach et al. introduced a novel compression technique based on folding-net, which projects 3D point clouds onto 2D grids and vice-versa, thereby enabling the use of traditional 2D image codecs for compression~\cite{9191180}. However, the folding process may introduce additional distortion, resulting in unsatisfactory compression performance.

The success of sparse convolution in compressing point cloud geometry has led to its application for attribute compression as well. Wang et al. modeled the point clouds as sparse tensors and processed them by a series of stacked sparse convolution layers, termed SparsePCAC, which significantly narrowed the performance gap with the traditional G-PCC~\cite{sparseW}.
Zhang et al. presented ScalablePCAC, a scalable compression technique for point cloud attributes~\cite{10313579}. ScalablePCAC integrates a standard G-PCC codec as the base layer and a sparse convolution-based model as the enhancement layer to handle full-resolution point cloud compression. An investigation of different quantization parameter combinations of the two layers is implemented to improve the rate allocation.
Further, Figueiredo et al. proposed an inherited embedded attribute encoding method by set partitioning in hierarchical trees and an MLP for context modeling~\cite{10474112}. Full scalability is achieved with a single code stream. 
\HXADD{Guo et al. proposed TSC-PCAC, an end-to-end voxel transformer and sparse convolution module (TSCM) based point cloud attribute compression~\cite{10693649}. TSCM extracts global and local inter-point relevance to enhance feature representation. Recently, Joint Photographic Experts Group (JPEG) published the deep-learning based coding architecture, modules, and the deep learning models involved in the encoding and decoding processes, which covers various coding configurations to achieve target compression rates.\cite{jpeg}}
In this paper, we focus on exploring the potential of end-to-end deep learning-based point cloud compression methods. Consequently, we compare the performance of our approach with the latest traditional point cloud compression method (i.e., G-PCC), the state-of-the-art deep learning-based methods (i.e., SparsePCAC) and a hybrid approach of traditional method and deep learning-based method (i.e., ScalablePCAC).

\subsection{Point Cloud-based Differentiable Rendering}
Point cloud rendering is an indispensable technique in the rendering of 3D objects. With the advent of differentiable rendering and neural rendering, it significantly contributes to various visual tasks across computer vision and graphics.

Insafutdinov and Dosovitskiy introduced a differentiable projection approach for point clouds, which allows for the efficient learning of 3D shapes from 2D projections~\cite{NEURIPS2018_4e8412ad}. This projection operator requires an explicit 3D volume to store point clouds, which limits the efficiency for large-scale point clouds.
Lin et al. proposed a generative modeling framework that efficiently generates 3D object shapes as dense point clouds~\cite{Lin_Kong_Lucey_2018}. A key component of their approach is the introduction of a pseudo-renderer, a differentiable module that approximates true rendering operations to synthesize novel depth maps.
DSS~\cite{10.1145/3355089.3356513} leverages the concept of surface splatting, projecting points as disks or ellipses onto the screen space, which is applied to point cloud geometry synthesis and denoising.

Novel view synthesis is a technique for generating new perspectives of point clouds, significantly benefiting from point cloud-based differentiable rendering.
Wiles et al. introduced SynSin, an end-to-end model for novel view synthesis using a differentiable neural point cloud renderer~\cite{Wiles_2020_CVPR}. It converts latent 3D feature clouds into new views, with synthesis quality potentially constrained by the accuracy of the learned 3D representations, which lack explicit geometric supervision.
You et al. created a locally unified 3D point cloud from sparse source views and utilized differentiable rendering to produce initial synthesized images, which were then enhanced to high quality through other jointly trained network components~\cite{10274683}.
Fridovich-Keil et al. proposed Plenoxels, a system for photorealistic view synthesis that represents a scene as a sparse 3D grid with spherical harmonics~\cite{Fridovich-Keil_2022_CVPR}. This approach optimizes the 3D grid from calibrated images without any neural components. 
Hu et al. presented TriVol, a 3D representation technique that uses slim volumes to create realistic images from point clouds~\cite{Hu_2023_CVPR}. It overcomes sparsity by merging features across scales, enabling view-consistent rendering without scene-specific fine-tuning.
Chang et al. presented Pointersect, which enables differentiable rendering of point clouds by directly inferring the intersection of a ray with the underlying surface represented by the point cloud~\cite{Chang_2023_CVPR}.
Kerbl et al. introduced a novel 3D Gaussian scene representation, coupled with the rasterization of splatting to achieve a real-time differentiable renderer~\cite{10.1145/3592433}. 

In this study, we adopt the standard point-based rendering~\cite{hearn2004computer,5980567} as the benchmark for the rendering quality,  and employ a differentiable manner~\cite{ravi2020pytorch3d,NEURIPS2019_bdbca288} to achieve the rendering-oriented point cloud compression task. The performance of new perspective images has also been evaluated, as depicted in Sec.~\ref{sec:com}.

\section{Proposed Method}
\label{sec:proposed method}
In this section, we systematically introduce the proposed rendering-oriented point cloud attribute compression. Sec.~\ref{sec:mot} explains the motivation of integrating rendering into compression framework, the use of sparse convolution and provides an overview of the framework. Sec.~\ref{com-mod} and Sec.~\ref{RM} detail the compression and rendering module, respectively. Sec.~\ref{loss} introduces the image rate-distortion loss function.

\subsection{Motivation and Overview}
\label{sec:mot}
\subsubsection{Impact of Rendering}
The integration of rendering within the compression framework is motivated by the fact that the direct reconstruction errors of the attributes are different from those of the rendered images through a typical rendering process. Taking the popular point-based rendering method~\cite{10.1145/1103900.1103907,rosenthal2008image,SAINZ2004869,7413904} as an example. The rendering process initiates with affine and projective transformations, which transform point clouds from 3D world coordinates into the 2D screen coordinates. The projective transformation alters the spatial relationships of the points, reflecting new geometric properties of the scene in the 2D context. Then the point clouds are rasterized, in which the points are modeled as circular regions. The attribute values of the centers in these regions correspond to the attributes of the transformed points. The attribute values of regions that lie at a distance from the center gradually decline as the distance increases. Moreover, the single pixel value on the screen is the weighted sum of the attributes of multiple nearest circular regions. Managing these geometry and attribute transformations is crucial for rendering an image that is visually coherent and faithful to the original three-dimensional scene. More details of the rendering can be found in Sec.~\ref{RM}.

\begin{figure}[htpb]
	    \centering
            \subcaptionbox{\parbox{3cm}{\vspace{-8pt}\centering Ground Truth \\ \textit{Loot}}}{
            \includegraphics[width=26mm]
	    {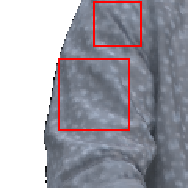}
            }
            \subcaptionbox{\parbox{3cm}{\vspace{-7pt}\centering reconstruction-\\oriented \\ bitrate: 0.0580 bpp}}{
            \includegraphics[width=26mm]
	    {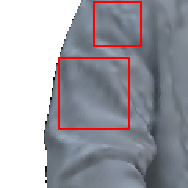}
            }
            \subcaptionbox{\parbox{3cm}{\vspace{-7pt}\centering rendering-\\oriented \\ bitrate: 0.0589 bpp}}{
            \includegraphics[width=26mm]
	    {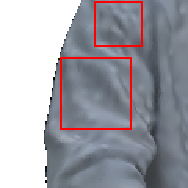}
            }
\vspace{10pt}
            \caption{Visual comparison of reconstructed point clouds of \textit{Loot} in similar bitrates. (b) is the proposed RO-PCAC without rendering module, which is trained by reconstruction loss. (c) is the proposed fully rendering-oriented RO-PCAC.}\label{moti}
\end{figure}

For compression methods aimed at fully reconstructing the input point cloud, the reconstructed point clouds with distortions are subjected to a secondary rendering process. The impact of rendering on distortion is difficult to quantify and consider. In contrast, our framework takes the influence of rendering into account in a differentiable manner, thereby establishing a novel paradigm for achieving optimal rendered image quality. As illustrated in Fig.~\ref{moti}, the rendered images of these two methods exhibit differences in their perception of visual depth. The results of the rendering-oriented framework are more true to the ground truth, owing to the inclusion of depth processing in the rendering module during training. 

\subsubsection{Significance of Sparse Convolution}
Compared with 3D dense convolution that processes every voxel uniformly in the 3D grid, sparse convolution only operates on predefined voxel sites on the sub-manifold, maintaining point cloud sparsity across layers and overcoming the dilation issue of the feature map.
The basic data structure is efficiently hash-indexed sparse tensor, which is characterized by two components: a coordinate set  $\vec{\mathbf{C}}=\left\{\left(x_i, y_i, z_i\right)=\vec{c}_{i}\right\}_i$ that delineates the positions of voxels, and a corresponding set of feature vectors $\vec{\mathbf{F}}=\left\{\vec{f}_{i}\right\}_i$. 

Then, sparse convolution can be defined as follows:
\begin{equation}\label{sconv}
\vec f_{\vec{c}_{i}}^{{\rm{out}}} = \sum\limits_{m \in K^3\left( {\vec{c}_{i},{{\vec{\mathbf{C}}}^{in}}} \right)} {{W_m}} \vec f_{\vec{c}_{i} + m}^{in}\quad {for}\quad \vec{c}_{i} \in {\vec{\mathbf{C}}^{{out}}},
\end{equation}
where ${\vec{\mathbf{C}}}^{in}$ and ${\vec{\mathbf{C}}}^{out}$ are input and output coordinate sets, respectively. $\vec f_{\vec{c}_{i}}^{{in}}$ and $\vec f_{\vec{c}_{i}}^{{out}}$ are the input and output feature vectors at the coordinate ${\vec{c}_{i}}$, respectively. $K^{3}\left(\vec{c}_{i}, {\vec{\mathbf{C}}}^{in}\right)=\left\{m \mid \vec{c}_{i}+m \in {\vec{\mathbf{C}}}^{in}, m \in K^{3}\right\}$ defines valid cells in a 3D convolutional kernel with size of $K^3$ and centered at $\vec{c}_{i}$, where $\vec{c}_{i}$ plus offset $m$ is in ${\vec{\mathbf{C}}}^{in}$. The element $W_m$ is the corresponding cell weight. In this paper, we continue to employ the sparse tensor as the underlying data structure and build upon the merits of sparse convolution to advance a sparse tensor-based transformer architecture. A detailed description of the transformer is provided in Sec.~\ref{com-mod}.

\begin{figure*}[htbp]
	    \centering
    	\includegraphics[width=181mm]
	    {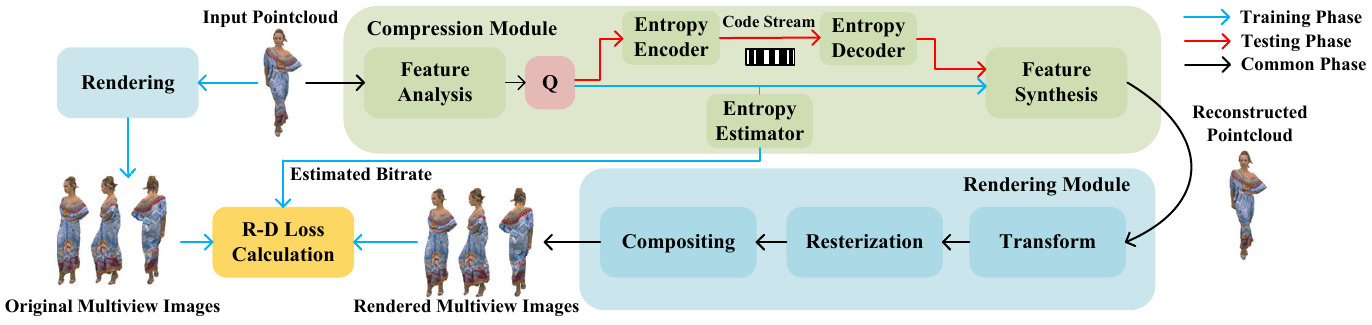}
        \caption{Overview of Rendering-Oriented Point Cloud Attribute Compression Framework. This framework comprises a compression module for feature analysis/synthesis, quantization (Q), and entropy encoding/decoding, and a rendering module for transforming, rasterization, and compositing of the point cloud data to multiview images. The rate-distortion (R-D) loss includes the estimated bitrate and the image error between original and rendered multiview images.} 
    	\label{systemview}
\end{figure*}

\subsubsection{Overview}
Fig.~\ref{systemview} illustrates the flowchart of the proposed rendering-oriented point cloud compression framework. The framework is divided into two primary modules: the compression module and the rendering module. The compression module efficiently compresses the input point clouds to compact code streams and decompresses them to reconstructed point clouds. \HXADD{Following the convention of point cloud attribute compression methods~\cite{g_pcc,sparseW,10693649,10313579}, we assume that the geometry has been losslessly transmitted.} The rendering module renders the reconstructed point clouds to images from preset multiple  perspectives through differentiable rendering techniques. Note that the rendering component of our framework is modular, allowing for the potential integration of future advancements in differentiable rendering techniques to replace the current rendering module used in this study.

\subsection{Compression Module}\label{com-mod}

\begin{figure}[tbp]
	\centering
	\subcaptionbox{Feature analysis}{
	\includegraphics[width=27mm]
	{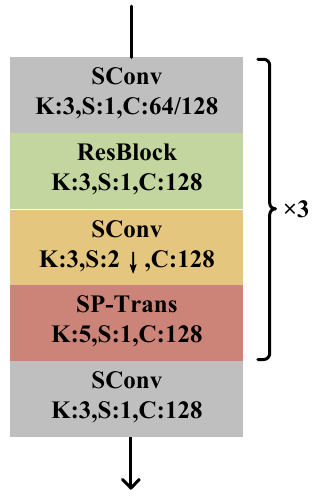}
	}
	\subcaptionbox{Feature synthesis}{
	\includegraphics[width=27mm]
	{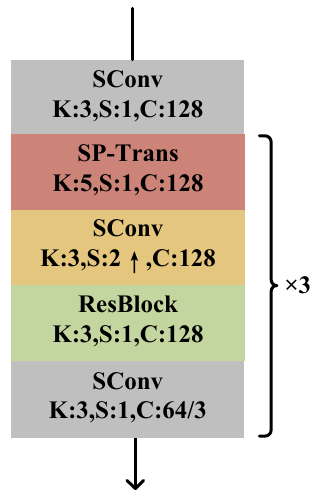}
	}
        \subcaptionbox{Residual block (ResBlock)}{
	\includegraphics[width=24mm]
	{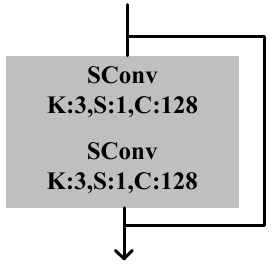}
	}
 \caption{The feature analysis and synthesis of the compression module. Except for the last convolution, each convolution is followed by a Rectified Linear Unit (ReLu) layer.}
 \label{syn1}
\end{figure}

\textbf{Feature analysis and synthesis.} The compression module is established on a variational auto-encoder structure~\cite{vae}, with sparse convolutions and sparse tensor-based transformers stacked in feature analysis and synthesis, as shown in Fig.~\ref{syn1}. The analysis transform starts at sparse convolution (SConv) layers with a stride of 1 and the output channel gradually increases to 64 and then to 128. Then, residual network layers (ResBlock) and downscaling sparse convolution with a stride of 2 are conducted to form more compact features from the attributes and reduce the number of points to be compressed. Finally, the sparse tensor-based transformers (SP-Trans) are followed to capture the dependencies between points and further enhance the features extracted. The structure of synthesis transform mirrors that of the analysis transform in reverse. 

\textbf{Sparse tensor-based Transformer.} Building upon the insights gleaned from the point-based transformers~\cite{Zhao_2021_ICCV,NEURIPS2022_d78ece66,guo2021pct}, we propose a novel architecture of sparse tensor-based transformer termed SP-Trans. This design is predicated on the innovative integration of the sparse tensor representation with the local self-attention mechanism. The SP-Trans is architected to address the inherent challenges in PCAC, where traditional convolutional operations struggle due to the sparse and irregular distribution of points in a point cloud. 

\begin{figure}[htbp]
	    \centering
    	\includegraphics[width=80mm]
	    {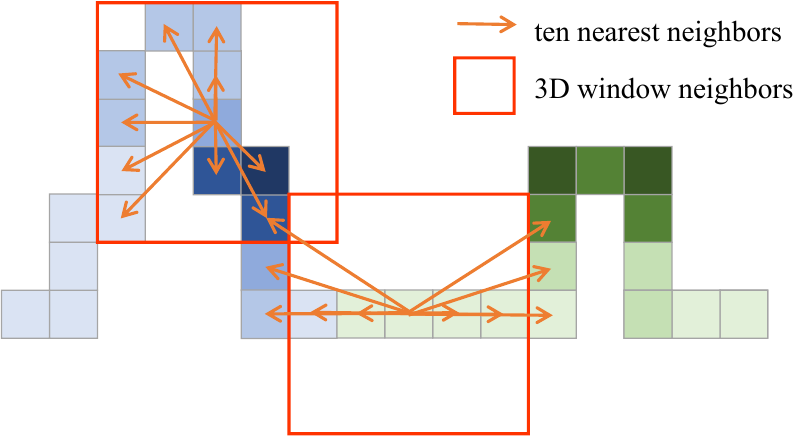}
        \caption{Two methods for searching nearest neighbors.} 
    	\label{SP-Trans}
\end{figure}

The core rationale behind SP-Trans is threefold. Firstly, sparse tensors offer an efficient data structure for managing the inherent sparsity in point clouds, which facilitates computation by focusing only on the occupied points. Secondly, as the nearest neighbors generally exhibit stronger correlations with the current point, SP-Trans employs an adaptive local neighborhood construction for each point, whereby neighbors are identified within a fixed-size 3D window. As illustrated in Fig.~\ref{SP-Trans}, neighboring points of two points are searched by ten nearest neighbors and a 3D window of fixed-size of 5, respectively.

In regions with higher point density, the window will naturally include more points, while in sparser regions, it will include fewer points. This adaptability ensures that the self-attention mechanism is always works with a consistent number of neighbors relative to the local point distribution. Conversely, employing a static number of neighbors, as in the k-nearest-neighbor-based (KNN) approach, may lead to the inclusion of a disproportionate number of points in sparse regions. This could result in considering points that are less relevant with respect to the local neighborhood, potentially diminishing the efficacy and precision of the self-attention mechanism. In addition, we can reuse the voxel-hashing in sparse convolution to query neighboring voxels in a 3D window with O(N) complexity, while KNN-based methods~\cite{Zhao_2021_ICCV,NEURIPS2022_d78ece66,guo2021pct} have the complexity of O(NlogN), which become burdensome for processing point clouds with millions of points. 

\begin{figure}[htbp]
	    \centering
    	\includegraphics[width=80mm]
	    {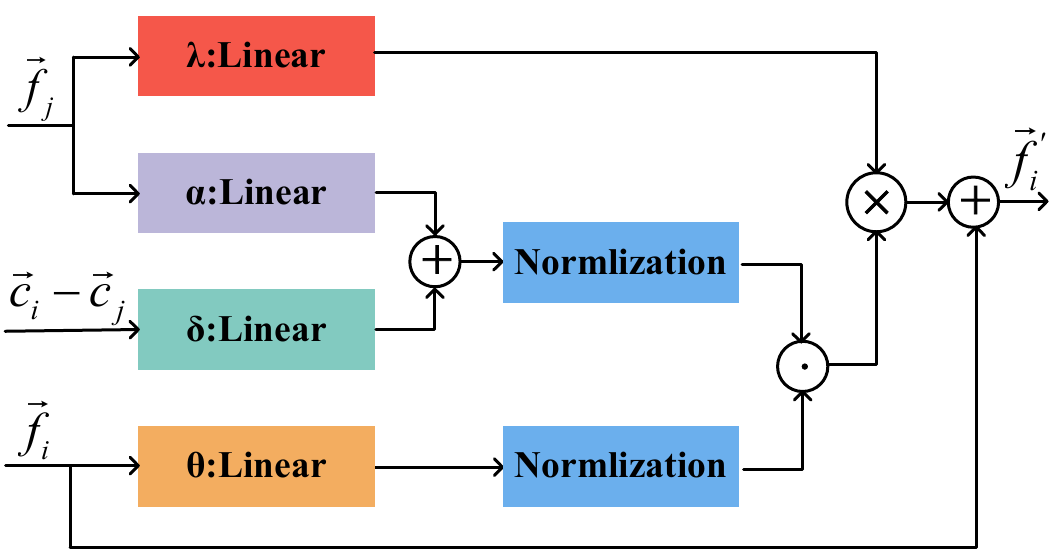}
        \caption{The architecture of SP-Trans.} 
    	\label{SP-Trans-overall}
\end{figure}

Thirdly, the local self-attention mechanism is adopted to capture long-range dependencies and the complex relationships between points. Given the local neighbor indices of $\vec{c}_{i}$ denoted by $\mathcal{N}(i)$, local self-attention on $\vec{c}_{i}$ can be formulated as follows:
\begin{equation}
\vec{f}_{i}^{\prime}=\sum_{j \in \mathcal{N}(i)} \left(\frac{\theta\left(\vec{f}_{i}\right)^\top \left(\alpha\left(\vec{f}_{j}\right) + \delta_{\mathrm{rel}}\left(\vec{c}_{i}-\vec{c}_{j}\right)\right)}{\left\|\theta\left(\vec{f}_{i}\right)\right\|\left\|\alpha\left(\vec{f}_{j}\right) + \delta_{\mathrm{rel}}\left(\vec{c}_{i}-\vec{c}_{j}\right)\right\|}\right)\lambda\left(\vec{f}_{j}\right)
\end{equation}
where $\vec{f}_{i}^{\prime}$ is the output feature and $\delta_{\mathrm{rel}}\left(\vec{c}_{i} - \vec{c}_{j}\right)$ is the relative position encoding. Functions $\theta$, $\alpha$ and $\lambda$ are the multi-layer perceptron-based query, key and value projection layers, respectively. The local self-attention layer utilizes the cosine similarity to measure the relationship between the $\theta\left(\vec{f}_{i}\right)$ and $\alpha\left(\vec{f}_{j}\right) + \delta_{\mathrm{rel}}\left(\vec{c}_{i}-\vec{c}_{j}\right)$ vectors, rather than the commonly used $\operatorname{softmax}(\cdot)$ function. This choice stems from the adaptive nature of the fixed-size window. Firstly, by focusing on the direction, it inherently handles similarity and makes a fair comparison when the number of neighbors varies. In contrast, the similarity calculated from the $\operatorname{softmax}(\cdot)$ function highly depends on the number of neighbors. Secondly, it prevents the potential issue of the $\operatorname{softmax}(\cdot)$ function. A single neighbor (i.e., $|\mathcal{N}(i)|=1$) would make the attention mechanism degenerate to a linear layer by normalizing the weight to unity. Thus, using cosine similarity in the local self-attention layer of SP-Trans is a proper choice that aligns with the adaptive, fixed-size window approach to neighbor searching. It ensures equitable weighting of neighbors that reflects the local point distribution. Combined with a skip connection, the final SP-Trans is illustrated in Fig.~\ref{SP-Trans-overall}.

\begin{figure}[htbp]
	    \centering
    	\includegraphics[width=88mm]
	    {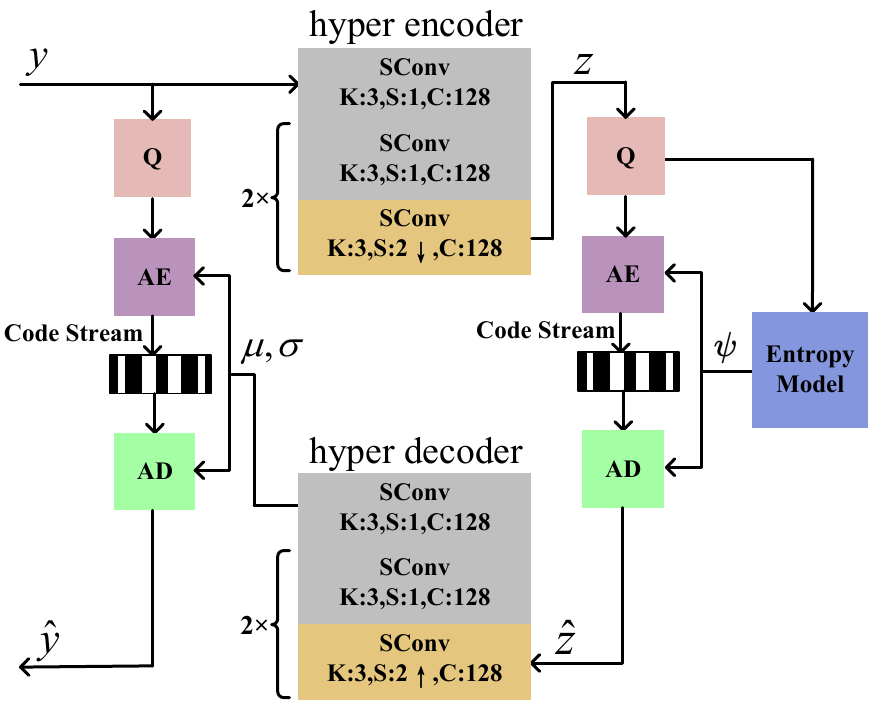}
        \caption{The architecture of the entropy model.} 
    	\label{em}
\end{figure}

\textbf{Entropy coding.} 
In the testing phase, the entropy coding is utilized to encode and decode the synthesized latent features $y$. After the quantization (Q) of $y$, the arithmetic encoder (AE) and arithmetic decoder (AD) are implemented for the respective processes of compression and decompression. To accurately model the distribution of $y$, we add hyperpriors~\cite{balle2018variational} to construct the entropy model as shown in Fig.~\ref{em}. The hyperencoder is tasked with generating a compact set of side information, represented as hyperpriors $z$. These hyperpriors are modeled by a fully factorized probability model~\cite{balle2016end} with learned parameters $\psi$. After quantization, compression and decompression, the reconstructed hyperpriors $\hat{z}$ are subsequently processed by the hyperdecoder to yield the mean and scale parameters $(\mu, \sigma)$ of the assumed conditional Gaussian distribution of $y$. 

In the training phase, the entropy estimator estimates the bitrate $R_{y}$ and $R_{z}$, which are respectively calculated by the entropy expectation of the latent and hyperprior features as:
\begin{equation}
p_{z \mid \psi}(z \mid \psi)=\prod_{i}\left(p_{z_{i} \mid \psi_i}\left(\psi_i\right) * \mathcal{U}\left(-\frac{1}{2}, \frac{1}{2}\right)\right)\left(z_{i}\right)
\end{equation}
and
\begin{equation}
p_{y \mid z}(y \mid z)=\prod_{i}\left(\mathcal{G}\left(\mu_{i}, \sigma_{i}\right) * \mathcal{U}\left(-\frac{1}{2}, \frac{1}{2}\right)\right)\left(y_{i}\right),
\end{equation}
where $\psi_i$ are the parameters of each univariate distribution $p_{z_i|\psi_i}$ and $\mathcal{U}\left(-\frac{1}{2}, \frac{1}{2}\right)$ means uniform distribution ranging from $-\frac{1}{2}$ to $\frac{1}{2}$ to simulate the quantization error. Conditioned on $z$, a Gaussian distribution is used to estimate the probability density function (p.d.f.) of $y$. The estimated mean and scale parameters $(\mu_i, \sigma_i)$ of each element $y_i$ are derived from $z$. Finally, the bit rates of $y$ and $z$ are estimated using
\begin{equation}
R_{y}=-\sum_{i} \log _{2}\left(p_{y_{i} \mid z_{i}}\left(y_{i} \mid z_{i}\right)\right),
\end{equation}
\begin{equation}
R_{z}=-\sum_{i} \log _{2}\left(p_{z_{i} \mid \psi_{i}}\left(z_{i} \mid \psi_{i}\right)\right).
\end{equation}

\subsection{Rendering Module}\label{RM}
The point-based rendering methods have garnered considerable popularity due to their real-time capabilities and effectiveness in computer graphics. In our framework, a classical point-based rendering method is implemented, which involves the transformation, rasterization, and compositing of 3D point clouds to generate 2D images. Notably, leveraging the autograd mechanism of deep learning libraries~\cite{ravi2020pytorch3d,NEURIPS2019_bdbca288}, all these steps are differentiable, thus allowing the gradients to be propagated from pixel features backward to point features. Below, we will introduce these steps respectively.

Transformation is the process of converting coordinates of points from the world space to the screen space of the display device, which involves a concatenated coordinate transformations. The first step involves transforming the points from the world coordinate system to the camera view coordinate system followed by:
\begin{equation}\label{RT}
X_{\text {view }}=R X_{\text {world }}+T,
\end{equation}
where $X_{\text {world}}$ are the coordinates of the point in the world coordinate system, $X_{\text {view }}$ are the coordinates of the point in the camera view coordinate system, $R$ is the rotation matrix and $T$ is the translation vector. The values of $R$ and $T$ are determined by the user-defined elevation, azimuth and the distance between the target point cloud and the camera. 

\begin{figure}[tbp]
\centering
\subcaptionbox{viewing frustum\label{vf}}
{
\includegraphics[width=40mm]
{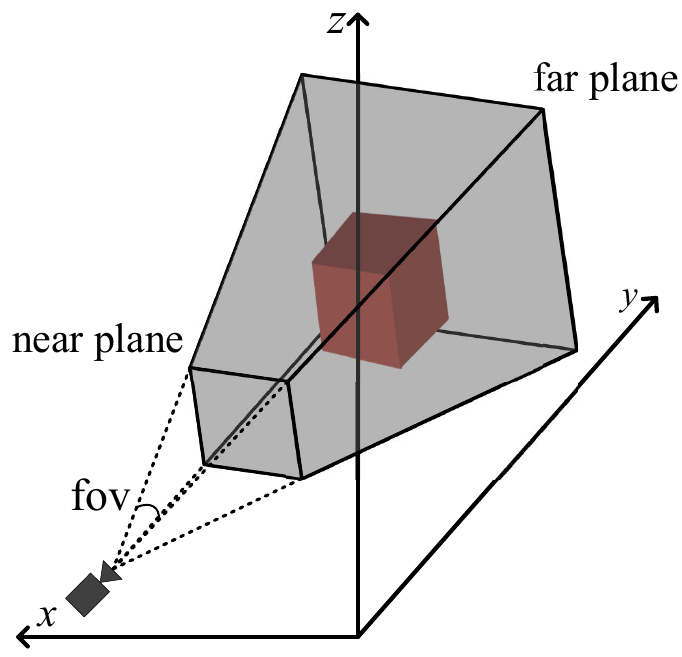}
}
\subcaptionbox{NDC system\label{Ns}}{
\includegraphics[width=40mm]
{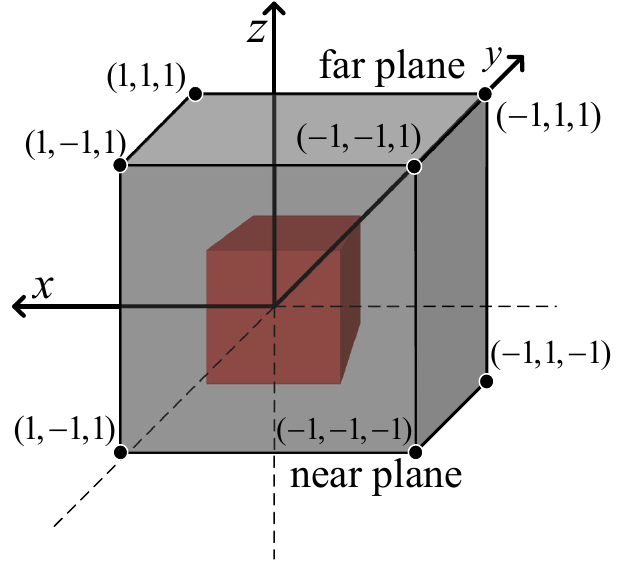}
}
\caption{The perspective transformation from camera space to NDC space}
\label{pers_trans}
\end{figure}

\begin{figure}[tbp]
	    \centering
        \includegraphics[width=50mm]
	    {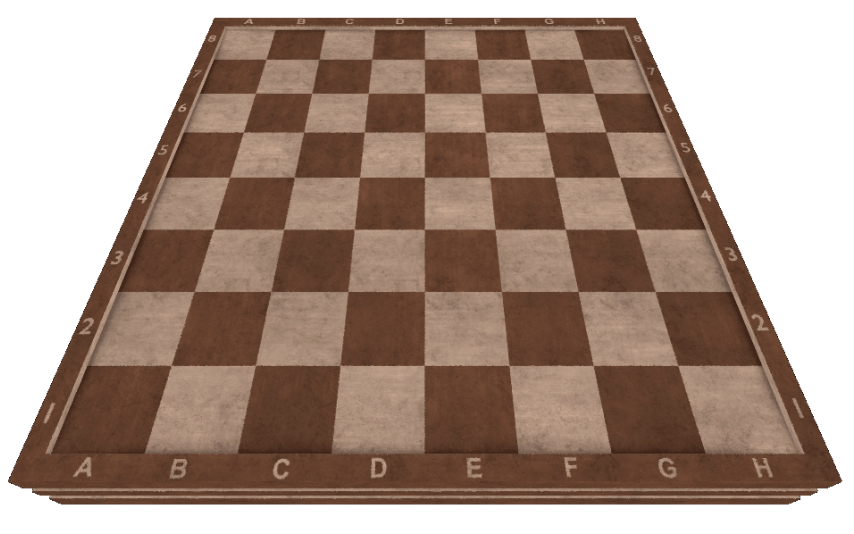}

    	\caption{The rendered image of the point cloud \emph{chessboard} generated by the rendering module.\label{chessboard}}
    	
\end{figure}

Once the points are in the camera view coordinate system, they need to be transformed into the normalized device coordinate (NDC) $X_{\text {ndc}}$. The specific type of camera (perspective or orthographic) will determine the projection matrix used for this transformation. The perspective camera, which more closely conforms to real scene shooting is chosen in our framework. 
\HXDEL{The coordinates are first placed into a symmetric viewing frustum as shown in Fig. 7(a) and the corresponding projective transformation transform the viewing frustum into a cubic space as shown in Fig. 7(b).}
\HXADD{The coordinates are first placed into a symmetric viewing frustum which represents the region camera shot, as shown in Fig.~\ref{vf}.} 
\HXDEL{The corresponding projective transformation, associated with a homogeneous division can be represented as}
\HXADD{Then the corresponding projective transformation~\cite{hearn2004computer}\footnote{\HXADD{This projective transformation refers equation (41) in Section \textit{Perspective Projections} of Chapter \textit{Three-Dimensional Viewing} on Page 334 of reference~\cite{hearn2004computer}. Due to minor differences in program implementation (i.e., left handed coordinate system vs right handed coordinate system and the range of mapped z coordinates [0,1] vs [-1,1] for more accurate depth precision), there are also minor differences in the z coordinate of NDC coordinates. However, due to the screen coordinates calculated from x and y components of NDC coordinates and the invariance of z (i.e., depth)-orders of points, this does not influence the effect of rendering, as indicated by Eq.~\eqref{weight} and Eq.~\eqref{color}.}}, associated with a perspective division can be represented as}
\setlength{\arraycolsep}{1.5pt} 
\begin{equation} \begin{aligned} \Tilde{X}_{\text{ndc}} &= \left[\begin{matrix} \frac{1}{\tan\left(\frac{fov}{2}\right) \times ap} & 0 & 0 & 0 \\ 0 & \frac{1}{\tan\left(\frac{fov}{2}\right)} & 0 & 0 \\ 0 & 0 & \frac{far}{far-near} & -\frac{far \times near}{far-near} \\ 0 & 0 & 1 & 0 \end{matrix}\right] \Tilde{X}_{\text{view}} \\ 
&\textcolor{black}{= \left[\begin{matrix} \frac{x}{ \tan\left(\frac{fov}{2}\right) \times ap} \\ \frac{y}{ \tan\left(\frac{fov}{2}\right)} \\ \frac{far\left(z-near\right)}{\left(far - near\right)} \\ z \end{matrix}\right]} \simeq \left[\begin{matrix} \frac{x}{z \times \tan\left(\frac{fov}{2}\right) \times ap} \\ \frac{y}{z \times \tan\left(\frac{fov}{2}\right)} \\ \frac{far\left(z-near\right)}{z\left(far - near\right)} \\ 1 \end{matrix}\right], \end{aligned} \end{equation}
where \HXADD{$fov$, $far$ and $near$ are the parameters of the viewing frustum. Specifically,} $fov$ is the field of view angle, $near$ is the distance of the camera to the near clipping plane, $far$ is the distance to the far clipping plane and $ap$ is the aspect ratio of the image pixels. \HXADD{The transformation of points in camera view coordinate system to NDC coordinates need to be calculated in the perspective space, where the coordinates are represented as homogeneous coordinates.} $\sim$ means the homogeneous coordinates of corresponding terms. \HXDEL{, which are used in this calculation in the projective space.} $\left[x,y,z,1\right]^\top$ are the homogeneous coordinates of $\Tilde{X}_{\text {view}}$. \HXADD{$\left[\frac{x}{ \tan\left(\frac{fov}{2}\right) \times ap},\frac{y}{ \tan\left(\frac{fov}{2}\right)},\frac{far\left(z-near\right)}{\left(far - near\right)},z\right]^\top$ are the homogeneous coordinates of $\Tilde{X}_{\text {ndc}}$ after the projective transformation. Then perspective division actually transforms the viewing frustum in the camera system into the cubic space in the NDC system, as shown in Fig~\ref{Ns}, which divides each component by the homogeneous $w$-component (i.e., $z$). Finally, we can get the 3D coordinates of $X_{ndc}$, i.e., $\small \left[\frac{x}{z \times \tan\left(\frac{fov}{2}\right) \times ap},\frac{y}{z \times  \tan\left(\frac{fov}{2}\right)},\frac{far\left(z-near\right)}{z \times \left(far - near\right)}\right]^\top$.} $\simeq$ denotes the equality under the homogeneous coordinates. \HXADD{In the subsequent and final viewport transformation, the $x$ and $y$ components will be transformed to the 2D screen coordinates by linear transformations.}
From the coordinates of \HXDEL{$\Tilde{X}_{\text {ndc}}$} \HXADD{${X}_{\text {ndc}}$}, we can see that the \HXDEL{horizontal and vertical} \HXADD{$x$ and $y$} components are scaled by the $z$-component (i.e., depth value). As a result, objects that are closer to the viewer appear larger than those that are farther away. This is how the projective transformation altered the geometry of the point cloud. As shown in Fig.~\ref{chessboard}, square grids of the chessboard will appear to converge and become trapezoidal in the distance. Without loss of generality, the farther an object is from the viewer, the more distorted its shape becomes. 

Finally, the viewport transformation, determined by the parameters of the rendering window such as its resolution, maps the \HXADD{first two dimensions of} the NDC coordinates, which range from $-1$ to $1$, to 2D screen coordinates. To provide a sense of depth and realism to the rendered 2D image, these sequential transformations present these points in 3D world with a new geometric relationship in 2D image.

\begin{figure}[htbp]
	    \centering
    	\includegraphics[width=60mm]
	    {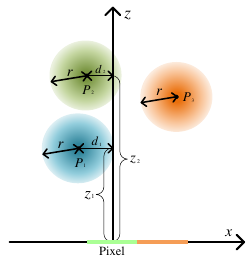}
        \caption{Rasterization of the adopted point-based rendering.} 
    	\label{ras}
\end{figure}

Subsequently, based on the altered geometry, the rasterization algorithm sorts the points in descending order of their depth values \HXDEL{$z$} \HXADD{in NDC system} to ensure that the closest points are processed first as shown in Fig.~\ref{ras}. This is achieved by calculating the depth value for each point and storing it in a z-buffer. Each point is splatted to a circular region in screen space, with opacity decreases away from the region’s center. The value of each pixel is computed by blending information from the K-nearest points in the z-buffer whose splatted regions overlap the pixel.

Compositing is the final stage responsible for blending the attributes of the points to generate the image. The compositor takes into account the color and alpha values associated with each point to determine the contribution of each point to the final pixel color. The process involves constructing weights $w_i$ for each point based on its distance to the pixel center, which is inversely proportional to the squared distance $d^2$ from the point to the pixel center, as shown in Eq.~\eqref{weight}.
\begin{equation}\label{weight}
w_{i}=1-\frac{d^{2}}{r^{2}}.
\end{equation}
Here, $r$ is the radius of the circular region used for rasterization, and $d$ is the Euclidean distance between the point and the pixel center. The final color $C$ of a pixel is computed by summing the weighted colors of all contributing points and normalizing by the total weight, as shown in Eq.~\eqref{color}. 

\begin{equation}\label{color}
\mathbf{C} = \sum_{i=1}^{K}\left(w_{i} \prod_{j=1}^{i-1}\left(1-w_{j}\right)\right) A_{i},
\end{equation}
where $A_i$ is the color of the i-th point of K nearest points, and the sums are taken over all points contributing to the pixel. Compositing uses the depth ordering of points so that nearer points contribute more. We can see that based on the altered geometry, the assignment of a single pixel value considers the colors of multiple points in the point cloud and their distances to the pixel center.

\subsection{Loss Function}\label{loss}
The design of the system also incorporates a critical evaluation component: the image loss. This module compares the rendered multiview images $\hat I$ with the original multiview images $I$, providing a quantitative measure of the fidelity of rendering. By assessing the pixel-wise differences, the system can refine its parameters to minimize the loss, ensuring that the rendered images are as true to the originals as possible. 

In addition, we want to represent $y$ and $z$ with as few bits as possible while minimizing the distortion between $I$ and $\hat I$. This can be achieved by minimizing the joint Lagrangian cost $J = R_{y} + R_{z} + \lambda \cdot d\left(I, \hat I\right)$. Distortion $d\left(, \right)$ between $I$ and $\hat I$ is measured by the mean squared error (MSE) and the trade-off between the bitrate and the distortion is controlled by the factor $\lambda$. 

\section{Experiments}
\label{sec:exp}
This section describes the comprehensive experimental framework, encompassing a detailed exposition of the training and testing datasets, hyperparameters in model training and testing and the selection of anchors. Furthermore, a series of objective and subjective quality assessment experiments are executed to show the superiority of RO-PCAC.
\subsection{Experimental settings} 

\subsubsection{Datasets}
\begin{figure*}[htbp]
	    \centering
            \subcaptionbox{Some samples of the training dataset\label{training_dataset}}{
            \includegraphics[width=59mm]
	    {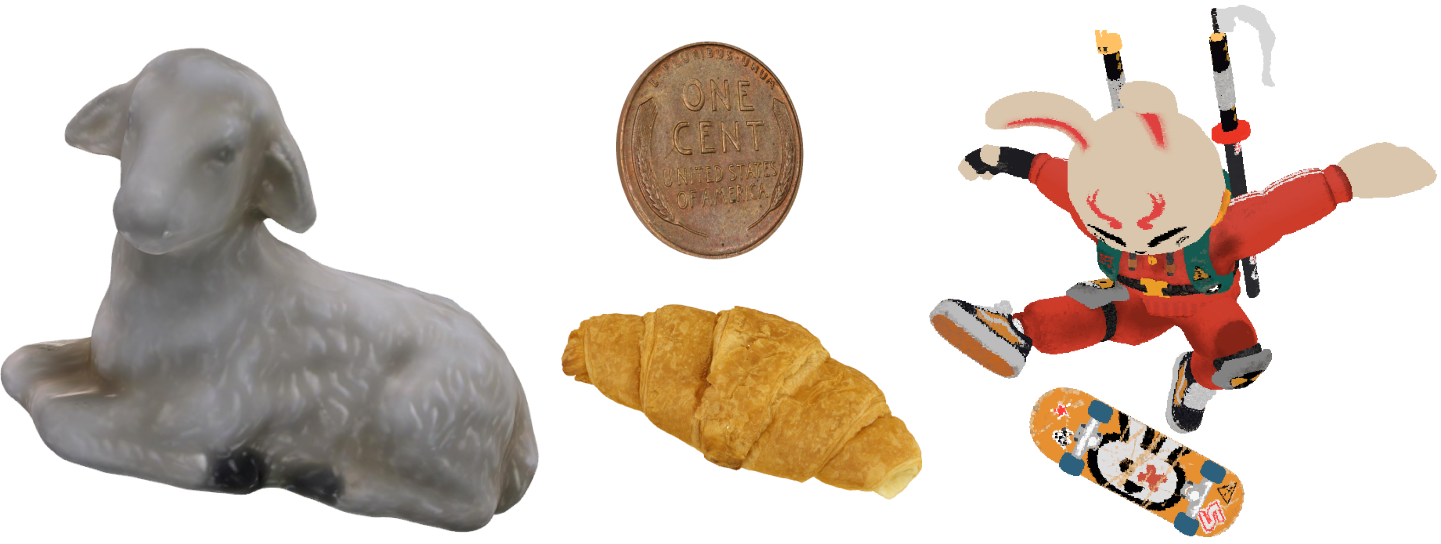}
            }
            \subcaptionbox{8iVFB\label{8iVFB}}{
            \includegraphics[width=52mm]
	    {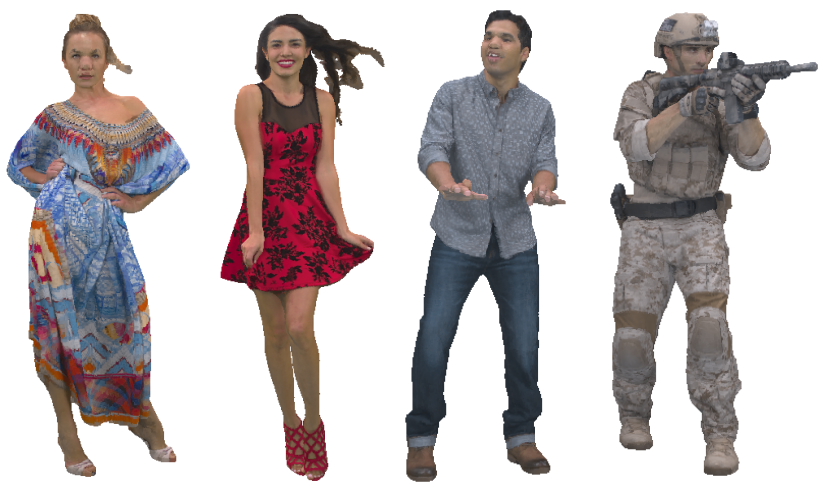}
            }
            \subcaptionbox{Owlii\label{Owlii}}{
            \includegraphics[width=60mm]
	    {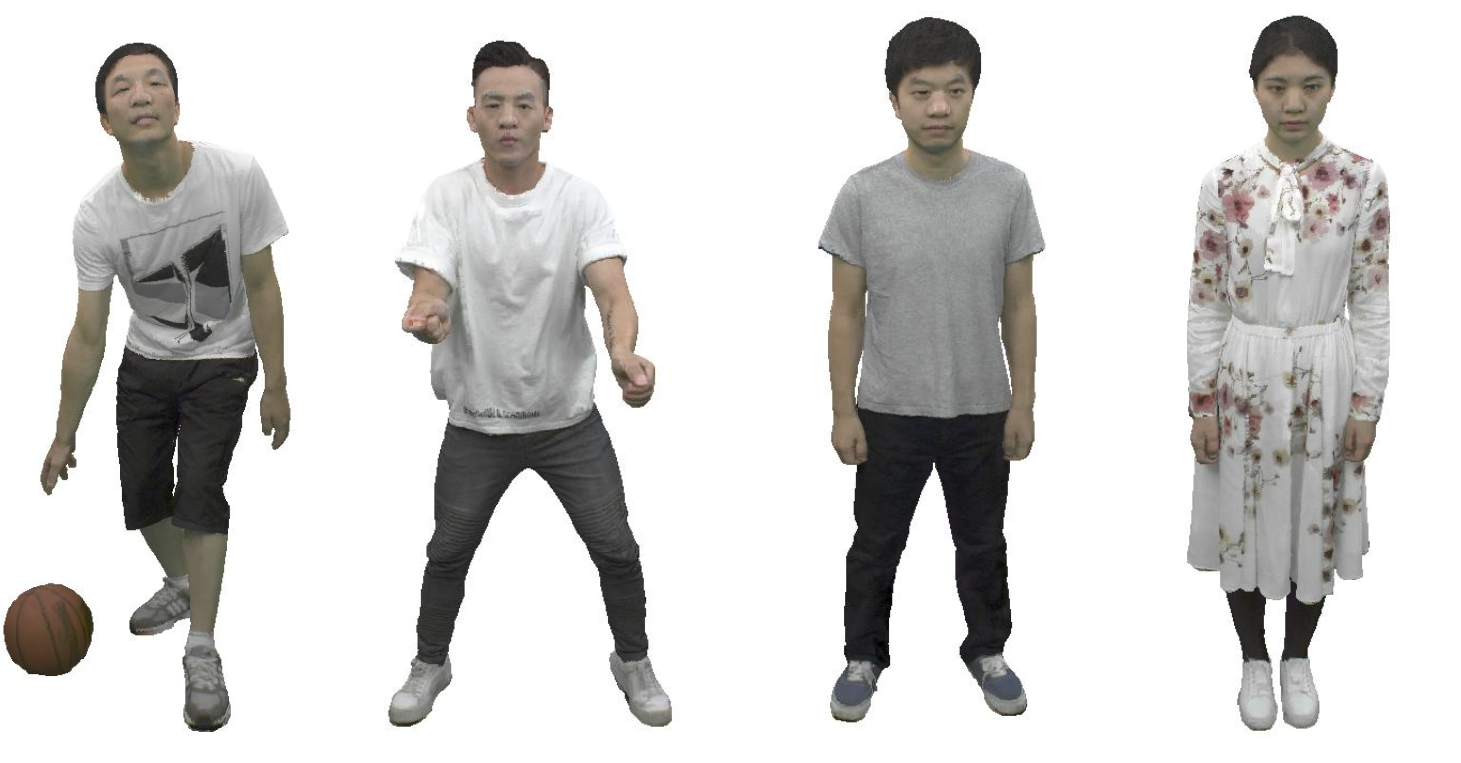}
            }
            \subcaptionbox{Some samples of ScanNet\label{scans_samples}}{
            \includegraphics[width=160mm]
	    {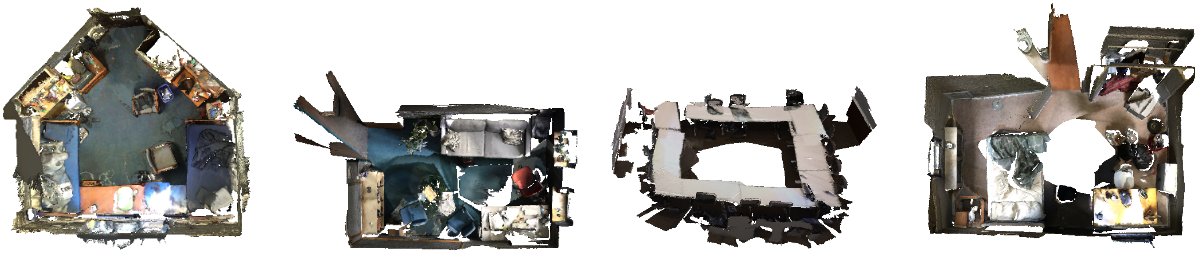}
            }
            \caption{Datasets.}\label{dataset}
\end{figure*}

\begin{itemize}
    \item Training datasets. The model is trained on 548 point clouds sampled from 3D models in the real-world textured things (RWTT) dataset~\cite{MAGGIORDOMO2020101943}. For each point cloud, $1 \times 10^6$ points are randomly sampled from the surface and then voxelized to a $1024^3$ space to maintain consistency with the test datasets. \HXADD{Due to the GPU memory limitation, we partition these 3D point clouds into 3D patches.} We show four training samples in Fig~\ref{training_dataset}.
    \item Testing datasets. Eight point clouds from two datasets used in MPEG are included in our testing samples. As shown in Fig.~\ref{8iVFB} and Fig.~\ref{Owlii}, four point clouds are defined in 8iVFB~\cite{8ivfb} and the other four are defined in Owlii~\cite{owlii}. \HXADD{ScanNet~\cite{dai2017scannet} is a large-scale indoor point cloud dataset containing 1603 scans at 2cm geometry precision. We used the officially divided training set of 1503 point clouds for training and the official test set of 100 point clouds for testing. Some samples in ScanNet are shown in Fig~\ref{scans_samples}.} The training and testing datasets are noticeably different, which demonstrates the generalization of our framework. 
\end{itemize}


\begin{figure}[htbp]
	    \centering
            \includegraphics[width=82mm]
	        {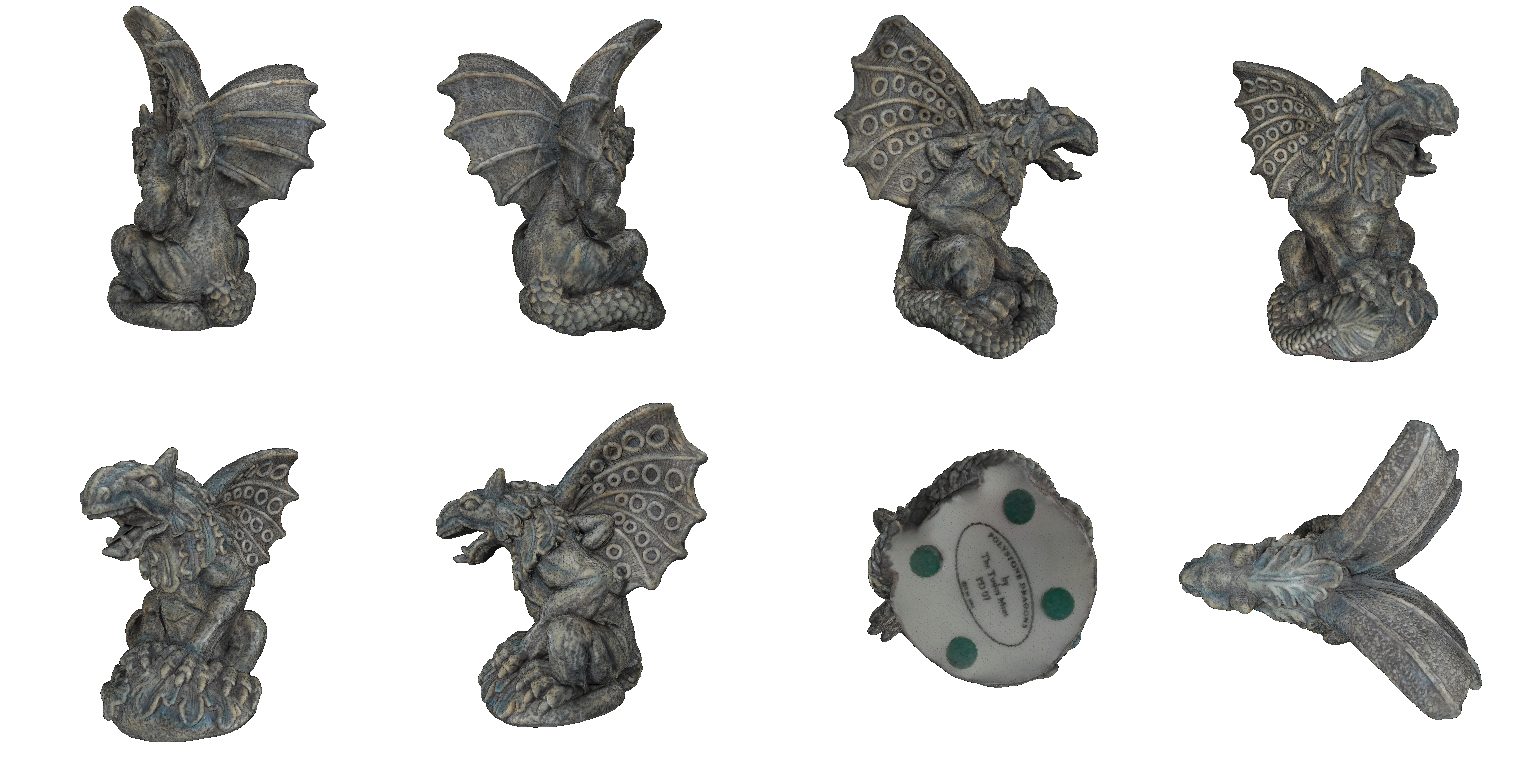}
            \caption{Multiviews of a training sample.}\label{multiviews}
\end{figure}

\subsubsection{Training and testing details}
\begin{itemize}
    \item Training phase. The ground truth multiviews include 8 views with elevations of $0^\circ$ and azimuths from $0^\circ$ to $360^\circ$ with an interval of $60^\circ$ and 2 views with elevations of $90^\circ$ and $270^\circ$ and azimuth of $0^\circ$, as shown in Fig.~\ref{multiviews}. \HXDEL{The resolution of each view's image is 384$\times$384.} \HXADD{To maximize the coverage of the training patches in the rendered images to emulate the effect of using 1024x1024x1024 images for training, we correspondingly reduced the training image resolution empirically to 384x384.} For the loss function, the value of $\lambda$ is set to 25,000, 4,000, 800, 250, and 85 for the training of the respective models. Each model undergoes training consisting of 140 epochs, facilitated by the PyTorch platform and the Adam optimizer. The parameters $\beta_1$ and $\beta_2$ of the Adam optimizer are configured at 0.9 and 0.999, respectively. The learning rate starts at 1e-4 and is progressively reduced to 1e-6, with a reduction to three-fourths of its previous value occurring every 15 epochs. The batch size is set to 8. The experimental setup includes a computer equipped with an Intel i9-14900k CPU operating at 3.60 GHz, 32 GB of main memory, and an NVIDIA GeForce RTX 4090 GPU.

    \item Testing phase. To demonstrate the generalization of RO-PCAC to different views, the ground truth multiviews include 6 views with the elevation of $0^\circ$ and azimuths from $30^\circ$ to $360^\circ$ with an interval of $60^\circ$. The resolution of each view's image is 1024$\times$1024\HXADD{, which matches the resolution of the test point cloud}. The R-D performance includes bpp for rate, PSNR and MS-SSIM \HXADD{of rendered images} for distortion. Furthermore, the BD-BR (Bjøntegaard Delta Rate)~\cite{bdrate} is used to evaluate the overall compression efficiency. The R-D performance in both Y and YUV (6:1:1) spaces~\cite{yuv611} is reported. 
\end{itemize} 
\subsubsection{Baselines}
\begin{itemize}
    \item G-PCC v14~\cite{g_pcc}. We use an improved version of RAHT with graph transform~\cite{9191183}, which is further enhanced by G-PCC v14 through entropy coding optimization. 
    \item G-PCC v23~\cite{g_pcc}. This is the latest point cloud compression method provided by MPEG, representing the state-of-the-art performance of traditional coding methods.
    \item SparsePCAC~\cite{sparseW}. This is the state-of-the-art end-to-end learning-based point cloud compression method. 
    \item \HXADD{TSC-PCAC~\cite{10693649}. This is an end-to-end voxel transformer-based point cloud compression method.}
    \item ScalablePCAC~\cite{10313579}. This is the state-of-the-art hybrid point cloud compression method with G-PCC as the base layer and a neural network as the enhancement layer.

\end{itemize}
\subsection{Compression Performance Comparison}\label{sec:com} 
\subsubsection{Objective Comparison.}

\begin{figure*}[htbp]
	    \centering
            \subcaptionbox{}{
            \includegraphics[width=40.6mm]
	    {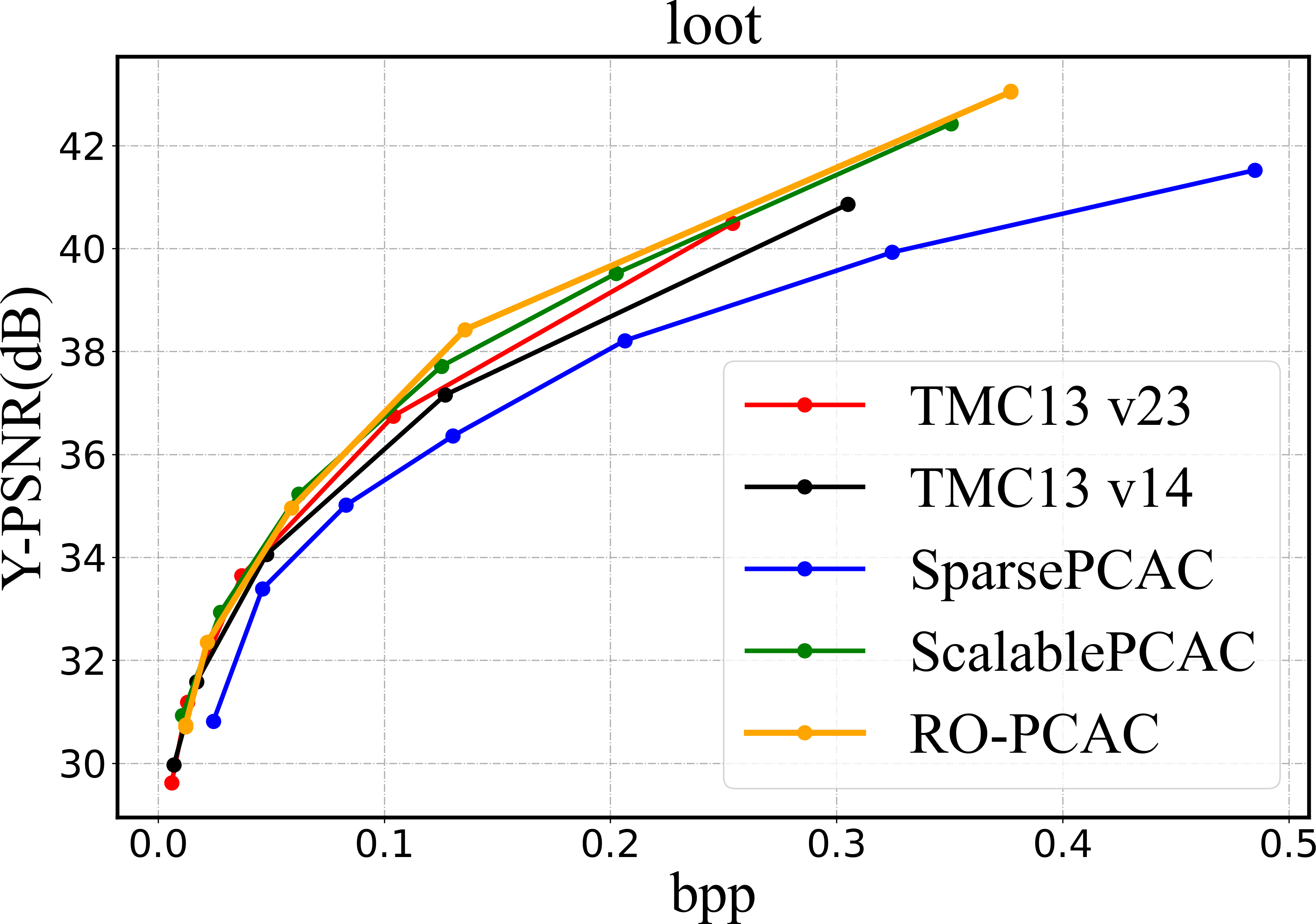}
            }
            \subcaptionbox{}{
            \includegraphics[width=43.5mm]
	    {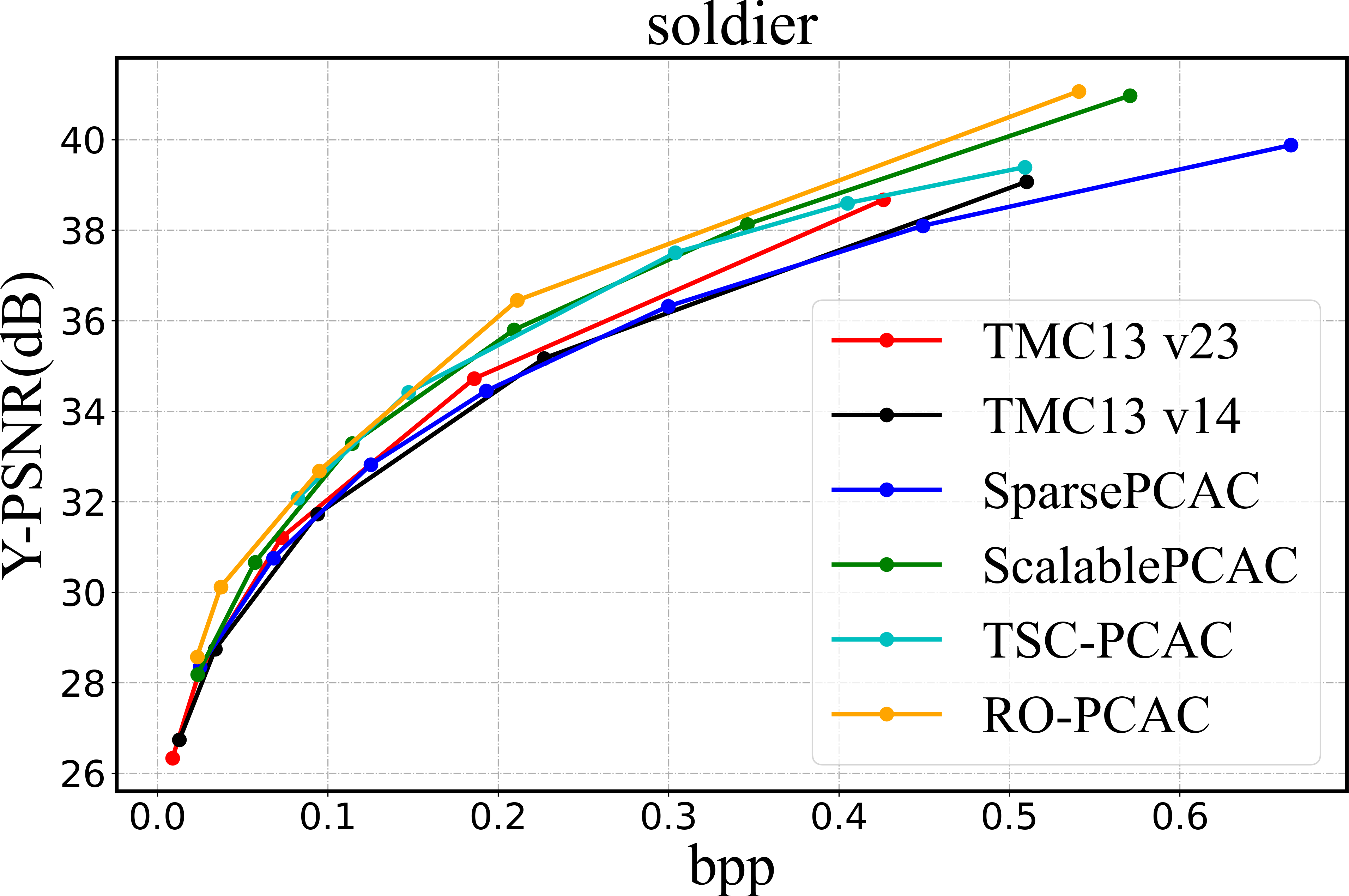}
            }
            \subcaptionbox{}{
            \includegraphics[width=40.6mm]
	    {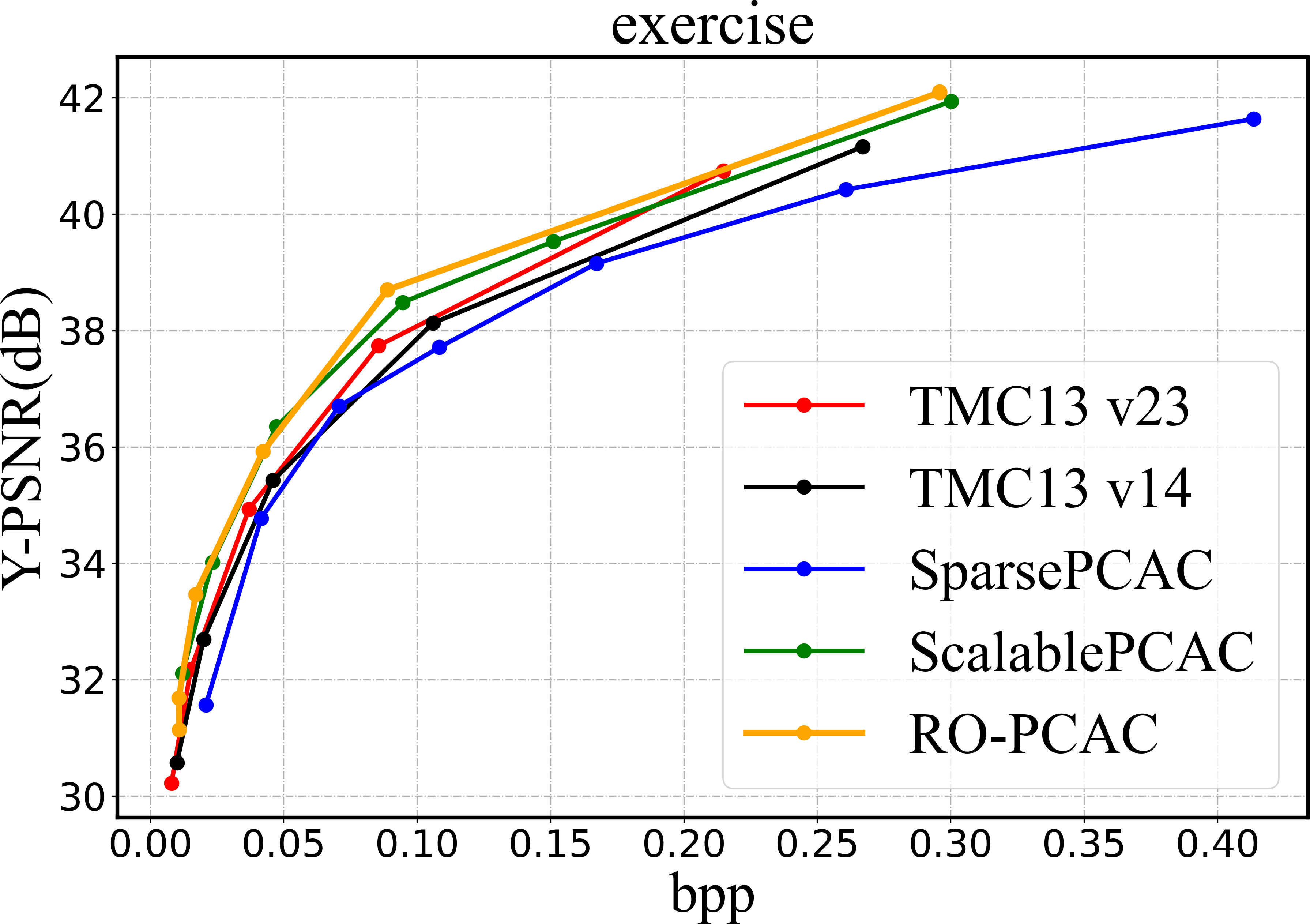}
            }
            \subcaptionbox{}{
            \includegraphics[width=43.5mm]
	    {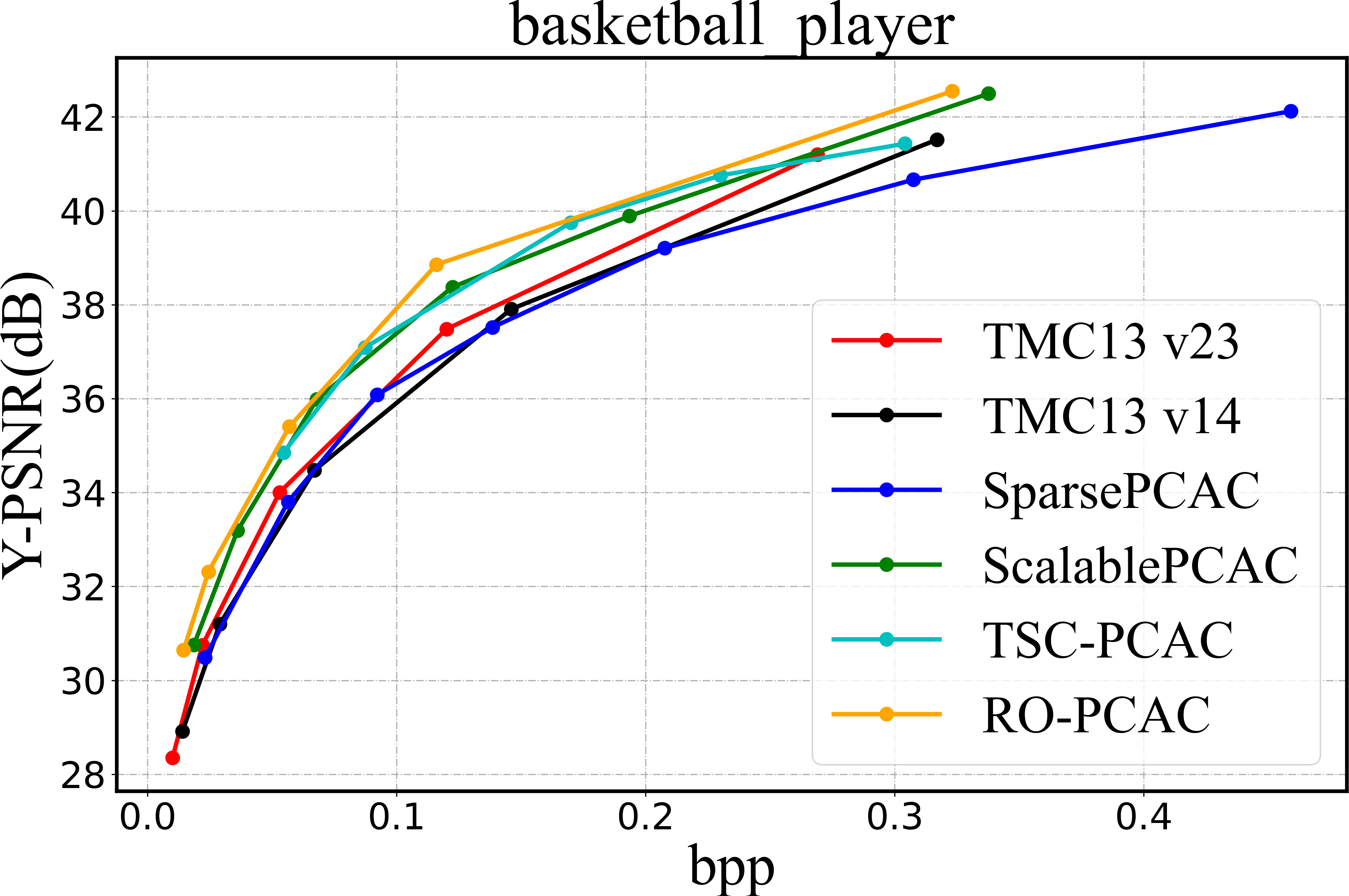}
            }

            \subcaptionbox{}{
                \includegraphics[width=40.6mm]
    	    {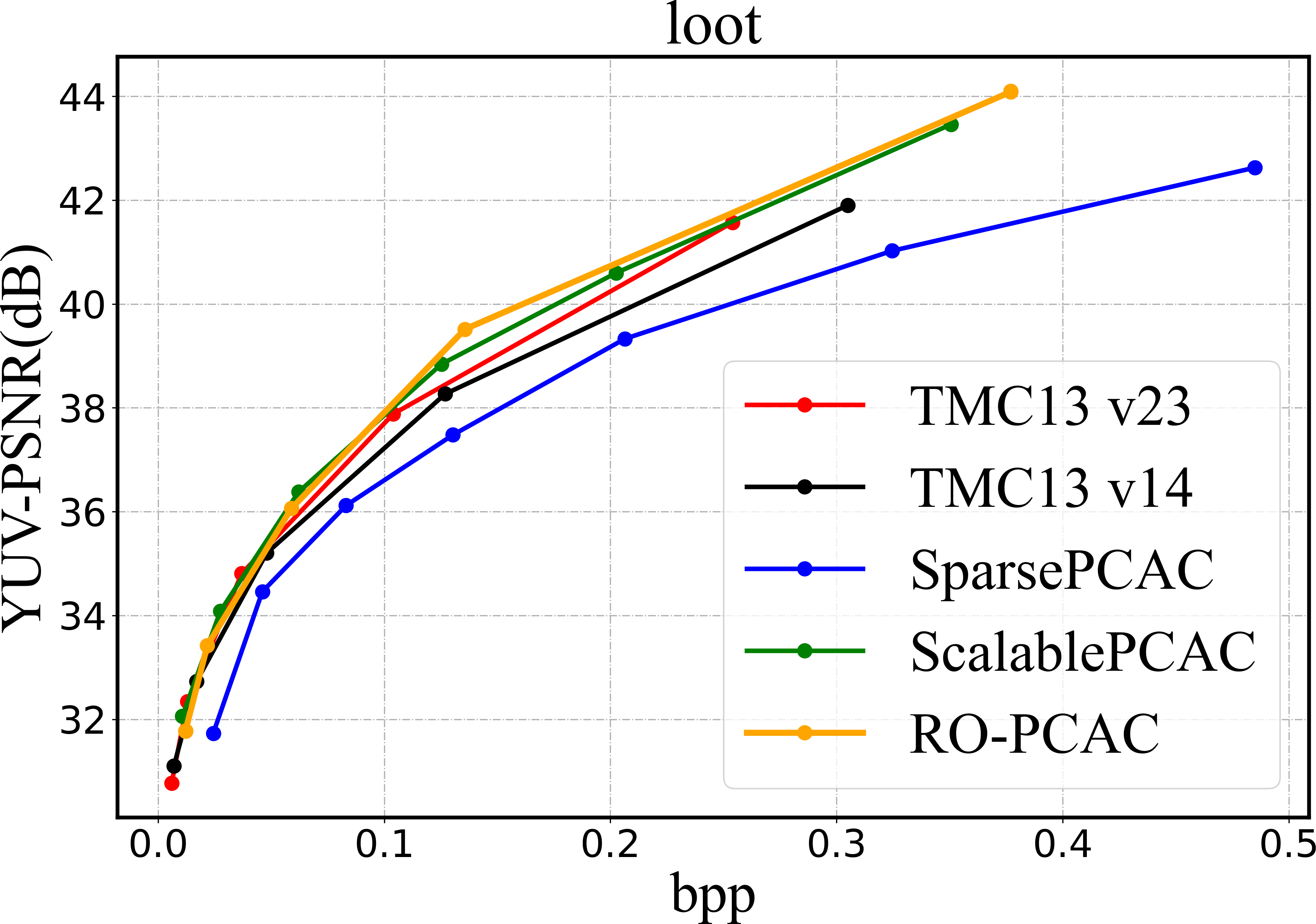}
                }
            \subcaptionbox{}{
            \includegraphics[width=43.5mm]
	    {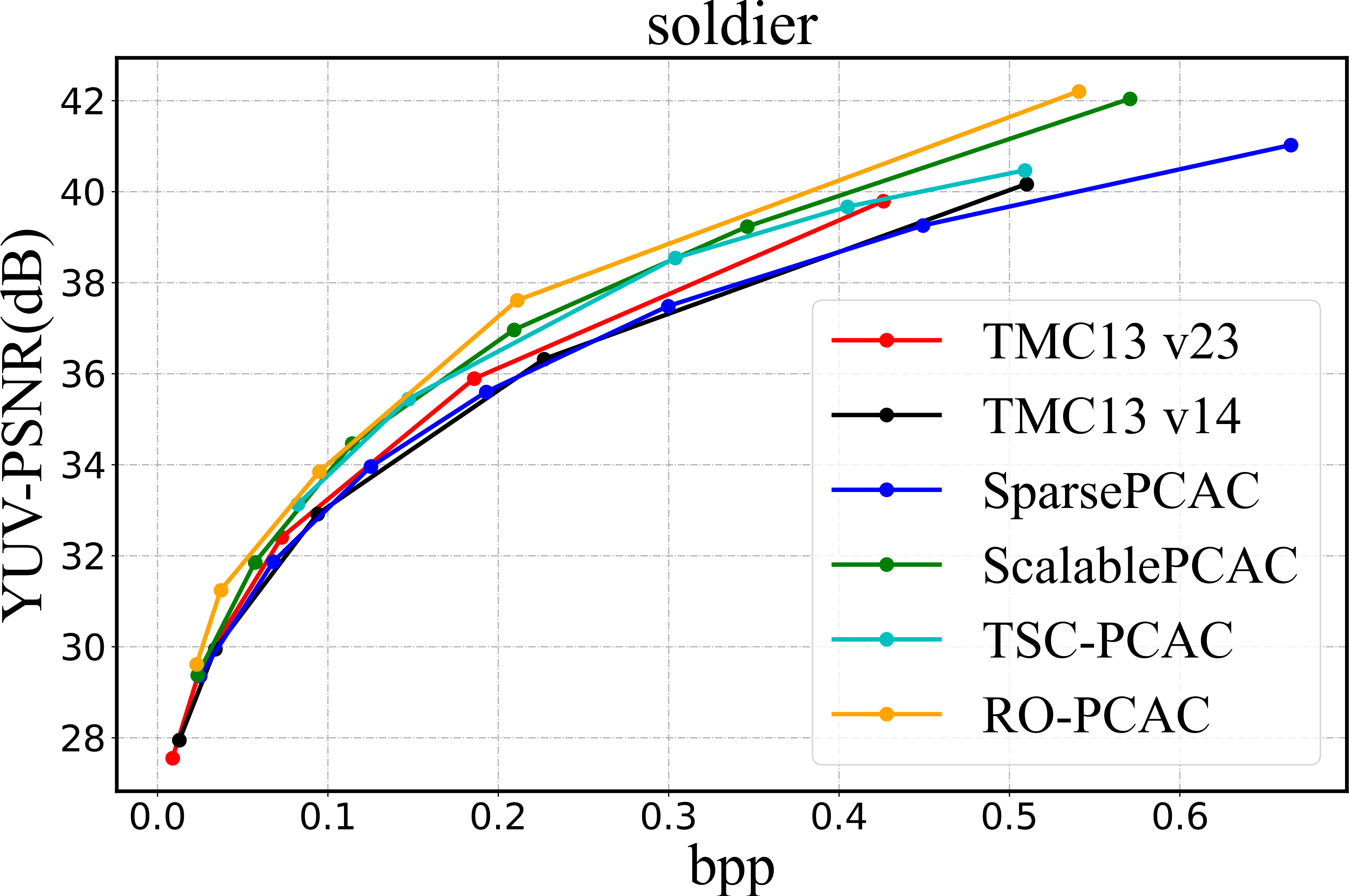}
            }
            \subcaptionbox{}{
            \includegraphics[width=40.6mm]
	    {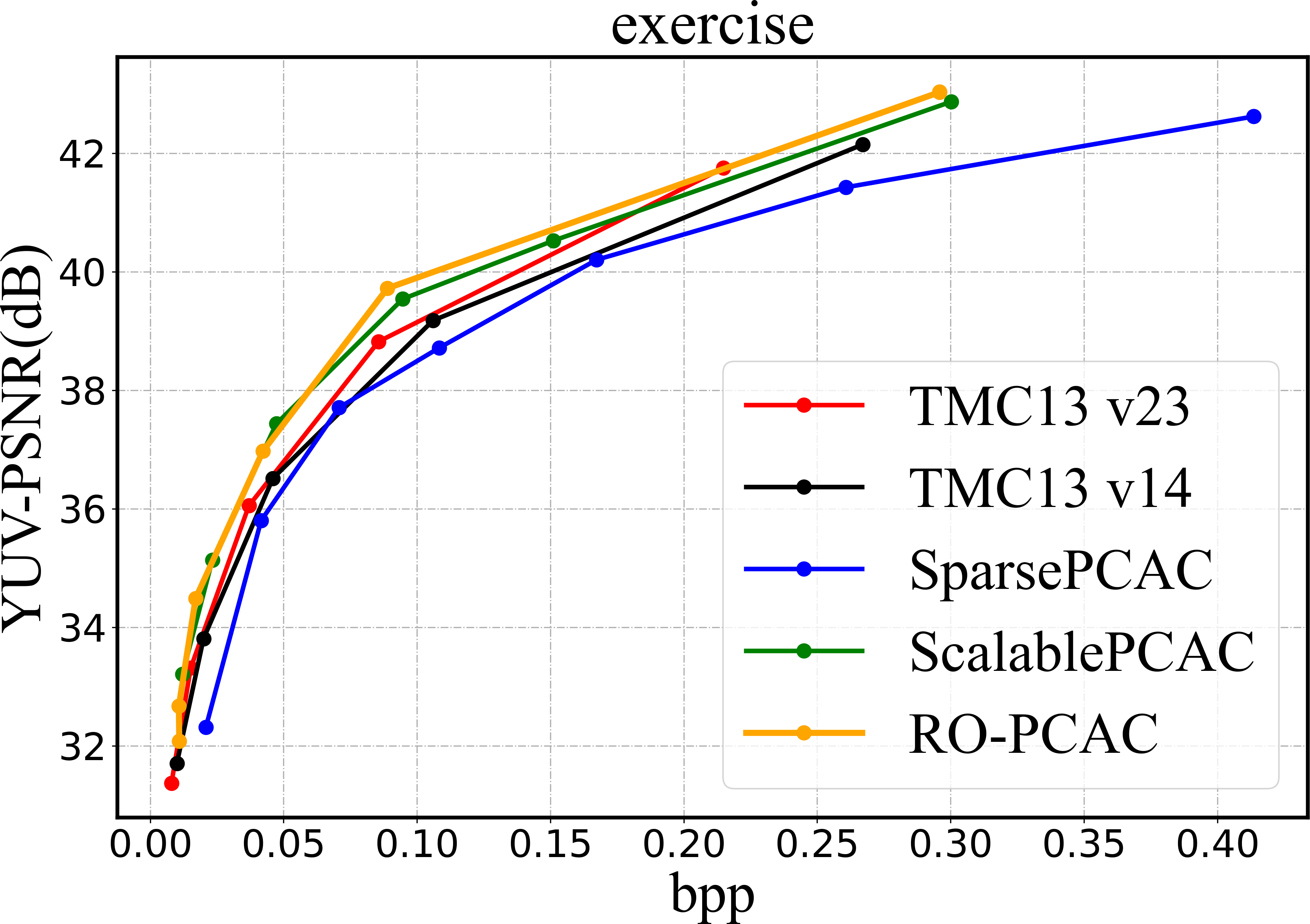}
            }
            \subcaptionbox{}{
            \includegraphics[width=43.5mm]
	    {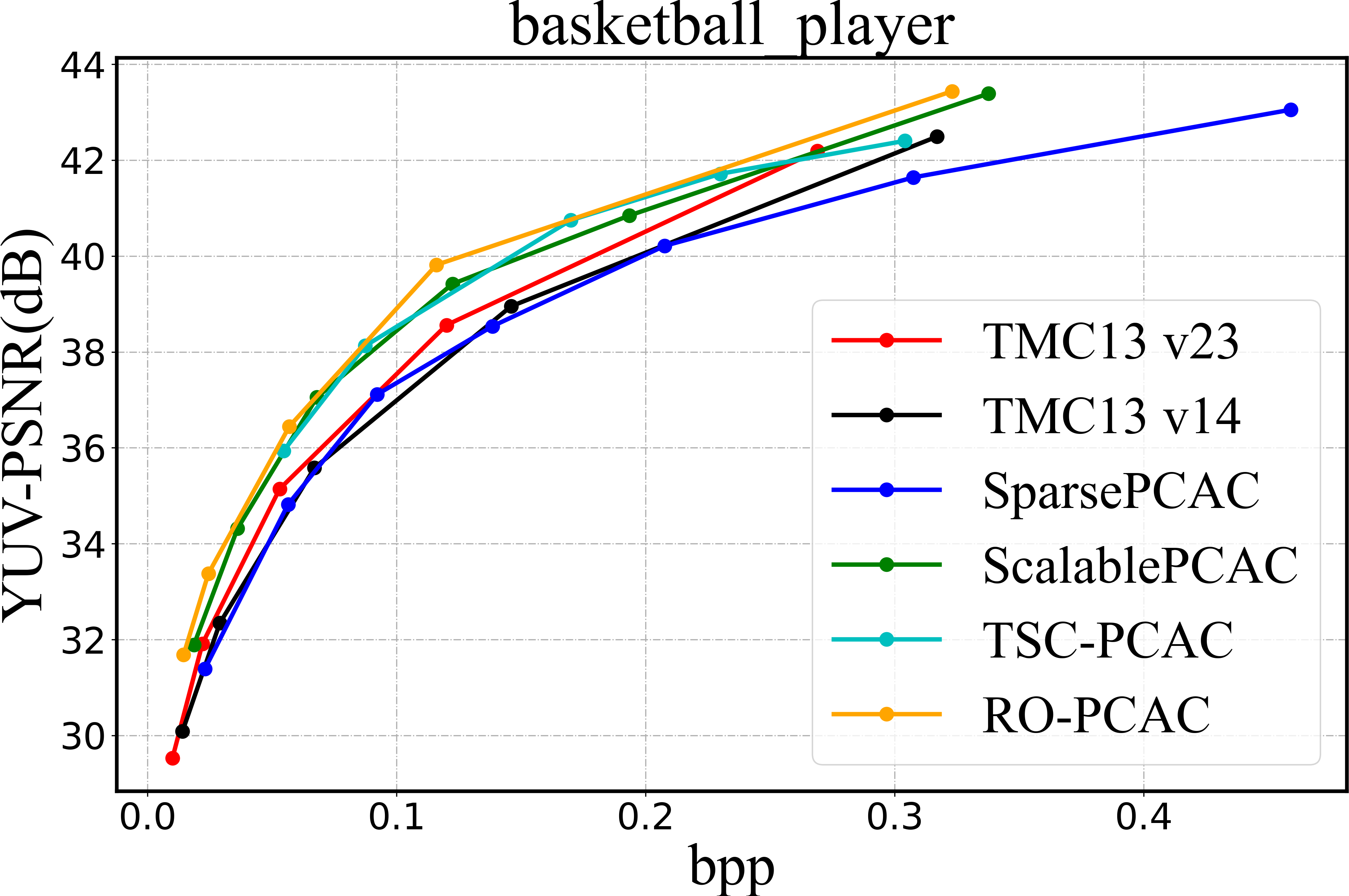}
            }

            \subcaptionbox{}{
            \includegraphics[width=43.4mm]
	    {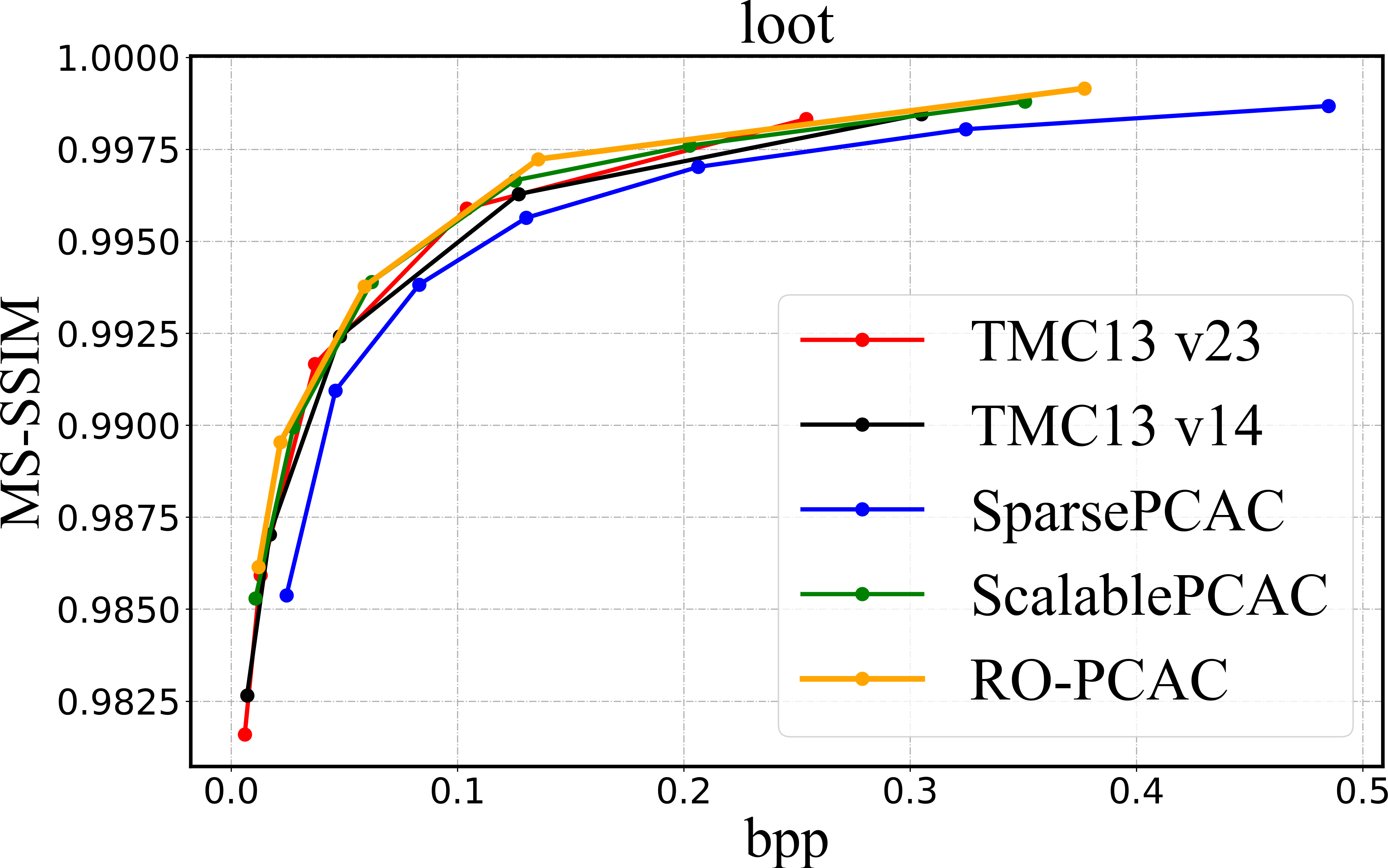}
            }
            \subcaptionbox{}{
            \includegraphics[width=42.5mm]
	    {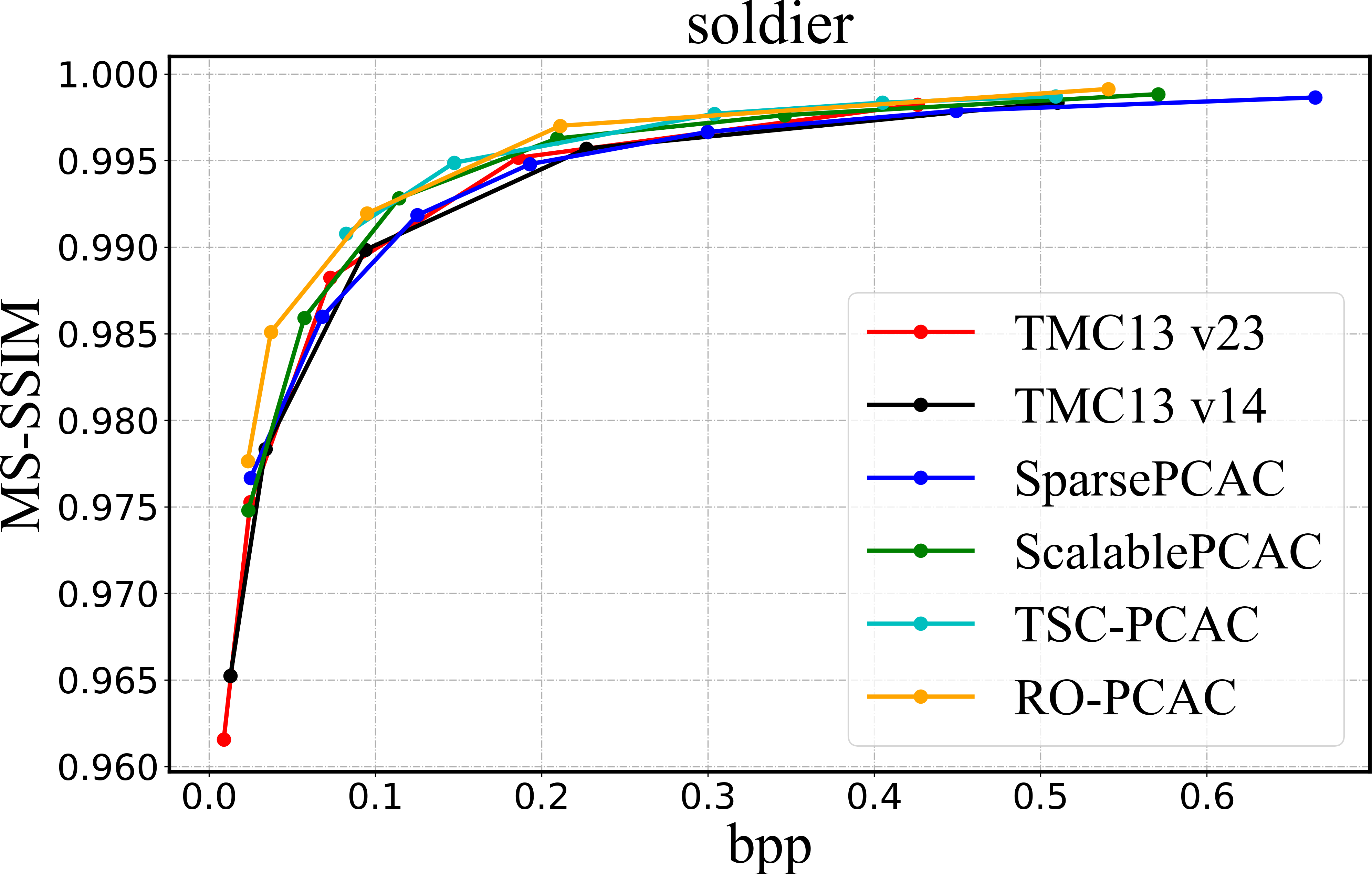}
            }
            \subcaptionbox{}{
            \includegraphics[width=39.8mm]
	    {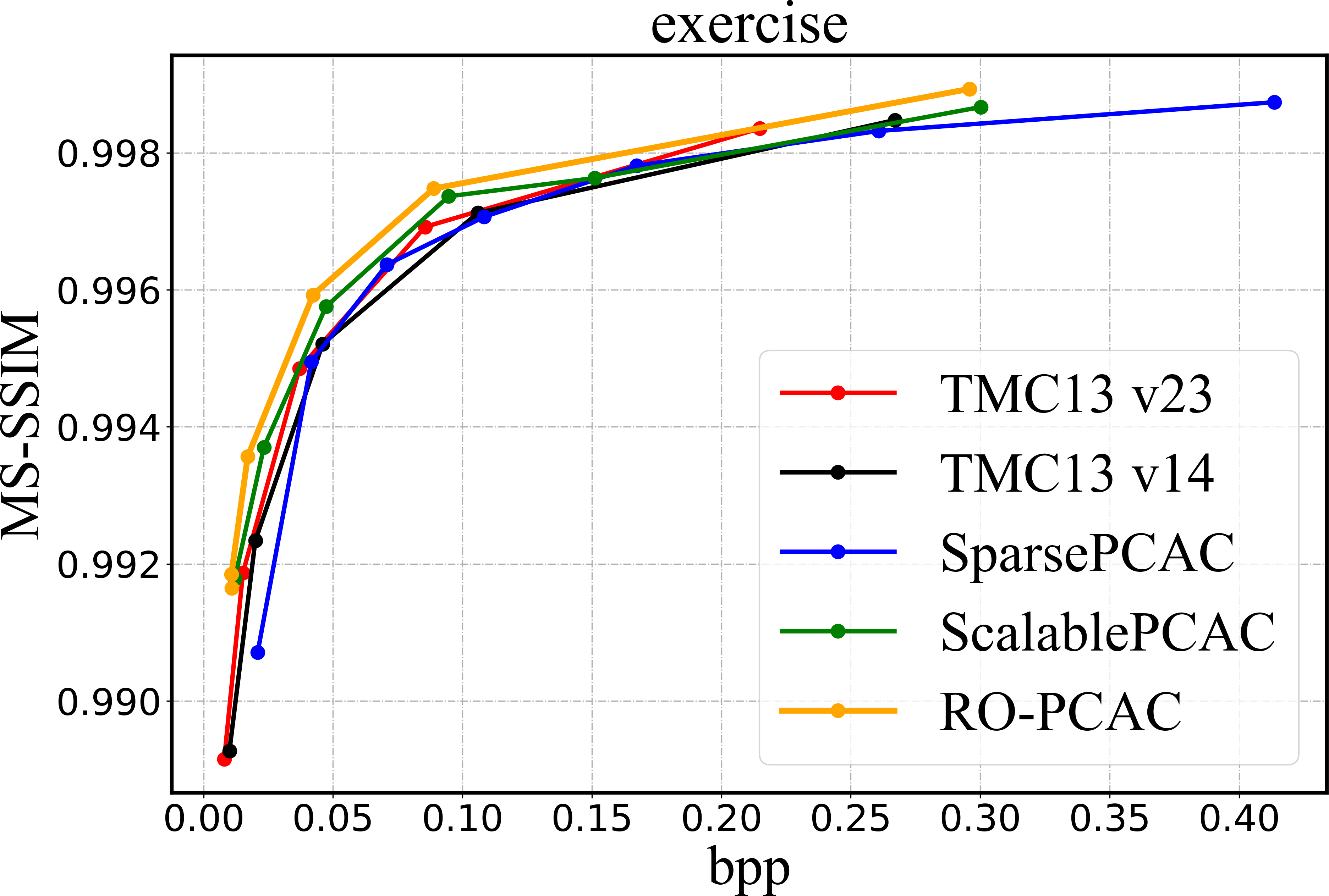}
            }
            \subcaptionbox{}{
            \includegraphics[width=42.5mm]
	    {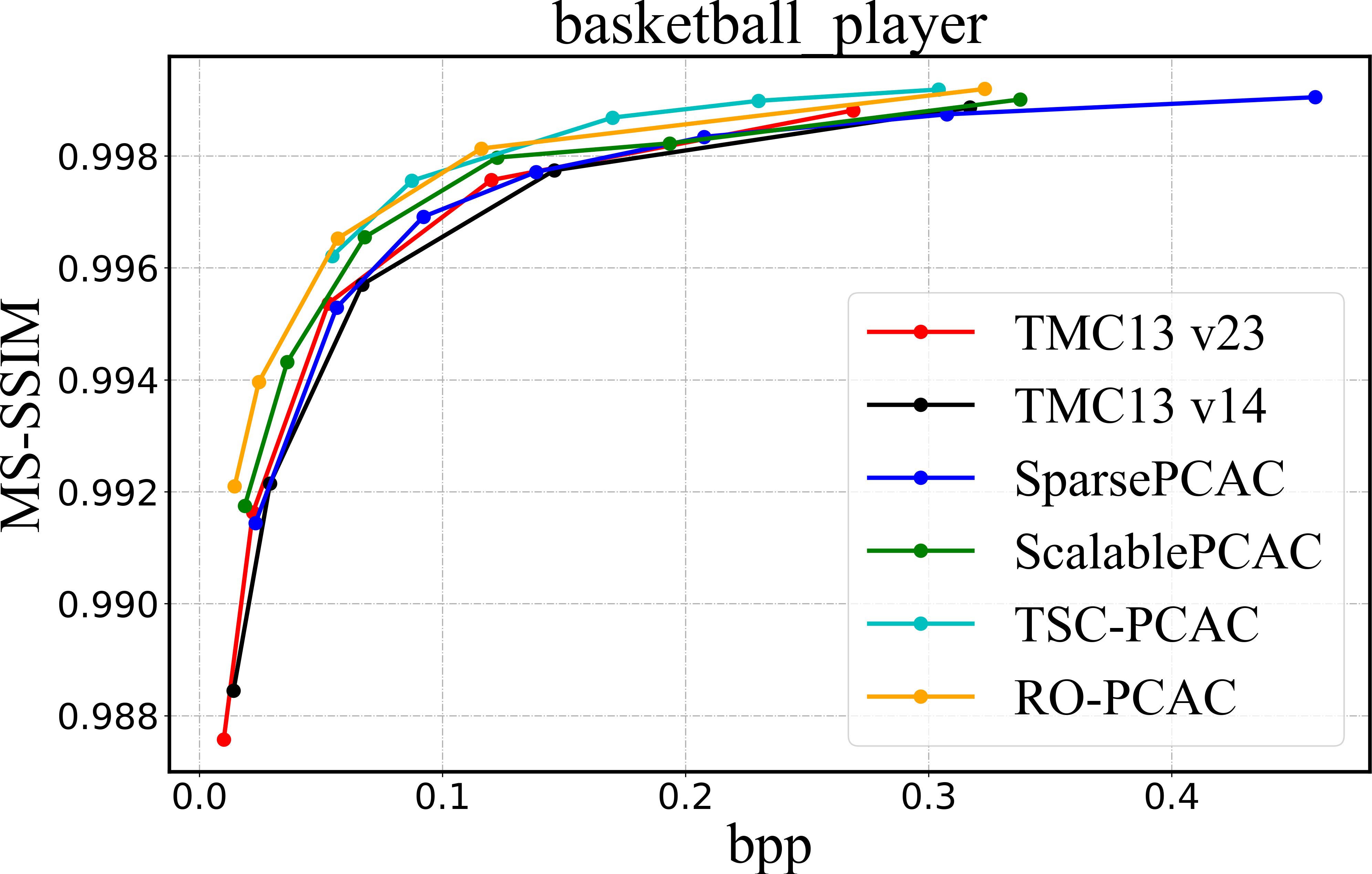}
            }
            
            \caption{R-D curve comparison with G-PCC, SparsePCAC and ScalablePCAC in the Y component, YUV components and MS-SSIM, respectively.}\label{curves}
\end{figure*}

\begin{table*}[htpb]
\centering
\caption{BD-BR (Y component, YUV components and MS-SSIM) comparison with TMC13 v14, TMC13 v23, SparsePCAC and ScalablePCAC}\label{yyuv}
\setlength{\tabcolsep}{2.5pt} 
\begin{tabular}{llccccccccccccccc}
\hline
\multirow{2}{*}{Class} & \multirow{2}{*}{Point Cloud} & \multicolumn{3}{c}{TMC13 v14~\cite{g_pcc}} & \multicolumn{3}{c}{TMC13 v23~\cite{g_pcc}} & \multicolumn{3}{c}{SparsePCAC~\cite{sparseW}} & \multicolumn{3}{c}{\textcolor{black}{TSC-PCAC}~\cite{10693649}} & \multicolumn{3}{c}{ScalablePCAC \cite{10313579}} \\ 
\cline{3-17}  &   & Y     & YUV     & MS-SSIM  & Y      & YUV   & MS-SSIM  & Y & YUV & MS-SSIM  & \textcolor{black}{Y}  & \textcolor{black}{YUV}  & \textcolor{black}{MS-SSIM} & Y  & YUV  & MS-SSIM \\ \hline
\multirow{5}{*}{8iVFB} & longdress          & -20.04& -14.24  &-13.49    &-18.24  &-10.40  &-5.05  & -10.65 & -10.25 & -10.38 &\textcolor{black}{\cmmnt{15.84}--}   &  \textcolor{black}{\cmmnt{31.95}--} &\textcolor{black}{\cmmnt{83.12}--}   &6.57   &4.56   &-8.22 \\  
                       & redandblack        & -27.81& -11.86  & 18.40   &-23.78  &-4.58   &30.57  & -23.05 &-30.48  & -37.17 &\textcolor{black}{-20.07}   &  \textcolor{black}{-18.23} &\textcolor{black}{-34.45}  &0.64   &-1.66  &-10.76 \\   
                       & loot               & -14.14& -12.92  &-16.85   &-5.84   &-3.93   &-9.40 & -36.70 &-36.88  &  -36.54 &\textcolor{black}{\cmmnt{3.25}--}   &  \textcolor{black}{\cmmnt{4.94}--} &\textcolor{black}{\cmmnt{21.64}--}  & -2.82 &-1.80  &-7.61 \\   
                       & soldier            & -27.90& -27.37  & -32.67  &-20.39  &-19.60  &-32.67 & -28.21 &-28.44  & -28.03 &\textcolor{black}{-10.06}   &  \textcolor{black}{-12.56} &\textcolor{black}{4.95}   & -11.35 & -10.92 & -16.17\\   \cline{2-17}
                       & \textbf{Average}   & \textbf{-22.66}& \textbf{-16.60}  & \textbf{-11.15}  &\textbf{-17.06}  &\textbf{-9.62}  &\textbf{-4.13}  & \textbf{-24.65} &\textbf{-26.51}  & \textbf{-28.03} &\textcolor{black}{\textbf{-15.07}}   &  \textcolor{black}{\textbf{-15.40}} &\textcolor{black}{\textbf{-14.75}}   &\textbf{-1.74}  &\textbf{-2.46}  &\textbf{-10.69}   \\   \hline
\multirow{5}{*}{Owlii} & basketball  & -33.62& -32.51  &  -41.09     &-25.81   &-24.17 &-33.34 & -35.17 &-35.21 & -31.36 &\textcolor{black}{-12.12}   &  \textcolor{black}{-11.35} &\textcolor{black}{-1.05}   &-12.05 &-10.71 &-20.15\\    
                       & dancer             &  -31.44& -30.51  & -36.66     &-24.85   &-23.54 &-30.54 & -29.04 & -29.43 & -31.27 &\textcolor{black}{-9.26}   &  \textcolor{black}{-9.44} &\textcolor{black}{-2.84}  & -9.21 &-8.50  &-15.45\\    
                       & exercise           &  -25.68& -24.46 & -34.26     &-19.19   & -17.04 &-27.17 & -36.80 & -37.55 & -33.13 &\textcolor{black}{\cmmnt{-1.75}--}   &  \textcolor{black}{\cmmnt{-0.46}--} &\textcolor{black}{\cmmnt{20.38}--}  & -8.43 &-7.33  &-19.34\\   
                       & model              & -28.66& -28.69  & -39.44     &-21.41   &-21.08 &-33.44 & -23.71 & -23.99 & -17.49 &\textcolor{black}{\cmmnt{-5.56}--}   &  \textcolor{black}{\cmmnt{-4.43}--} &\textcolor{black}{\cmmnt{12.01}--}  &-11.78 &-11.90 &-18.35\\   \cline{2-17}
                       & \textbf{Average}   & \textbf{-29.85}& \textbf{-29.04}  & \textbf{-37.86}     &\textbf{-22.82}   &\textbf{-21.46} &\textbf{-31.12} & \textbf{-31.18} & \textbf{-31.56} & \textbf{-28.31} &\textcolor{black}{\textbf{-10.69}}   &  \textcolor{black}{\textbf{-10.40}} &\textcolor{black}{\textbf{-1.95}}  &\textbf{-10.37}  &\textbf{-9.61}  &\textbf{-18.32} \\  \hline
\multicolumn{17}{l}{{‘}--{’} indicates the test point cloud is selected into training dataset.}\\
\end{tabular}
\end{table*}

Our method's superiority is underscored by the BD-BR reductions reported in Table~\ref{yyuv} and further visualized through the R-D curves in Fig.~\ref{curves}. Specifically, for the 8iVFB and Owlii datasets, our method outperforms G-PCC TMC13v14 by a significant margin, with a BD-BR reduction of 22.66\% (16.60\%) and 29.85\% (29.04\%) in Y (YUV) color space, respectively. Moreover, when compared to G-PCC TMC13v23, our method also achieves a notable BD-BR reduction of 17.06\% (9.62\%) and 22.82\% (21.46\%), respectively. 

Furthermore, our method was also compared with \HXADD{TSC-PCAC}, SparsePCAC and ScalablePCAC, \HXDEL{two} \HXADD{three} other learning-based models for point cloud attribute compression. To ensure a fair comparison, SparsePCAC was trained according to the training settings of this study. \HXADD{In the same way, we retrained TSC-PCAC using our training conditions, but the performance was not as good as the trained models provided by the authors. Therefore, we used their trained models to report their performance in our paper.} Due to the source code of ScalablePCAC not being open and the use of an identical training dataset in our study, we asked the author for the reconstructed point clouds at different ratepoints to compare. \HXDEL{Both} \HXADD{All} methods are tested under the same conditions. As shown in Table.~\ref{yyuv}, our method achieves an average BD-BR reduction of 24.65\% (26.51\%) and 31.18\% (31.56\%) over SparsePCAC in the two test datasets. \HXADD{Compared to TSC-PCAC, our method achieves a BD-BR reduction of 15.07\% (15.40\%) and 10.69\% (10.40\%), respectively.} Compared to ScalablePCAC, our method also achieves a BD-BR reduction of 1.74\% (2.46\%) and 10.37\% (9.61\%), respectively.

In addition, we have employed the MS-SSIM metric for a more comprehensive assessment of performance. Our method surpasses the G-PCC TMC13v14, TMC13v23, SparsePCAC, \HXADD{TSC-PCAC} and ScalablePCAC by 11.15\%, 4.13\%, 28.03\%, \HXADD{14.75\%} and 10.69\% BD-BR in 8iVFB, respectively. In the Owlii dataset, the respective improvements are 37.86\%, 31.18\%, 28.31\%, \HXADD{1.95\%} and 18.32\%.

\begin{table*}[htpb]
\centering
\caption{BD-BR (YUV components) comparison with TMC13 v14, TMC13 v23, SparsePCAC and ScalablePCAC in different resolutions}\label{resolutions}
\setlength{\tabcolsep}{4.5pt} 
\begin{tabular}{clccccccccccccccc}
\hline
\multirow{2}{*}{Class} & \multirow{2}{*}{Point Cloud} & \multicolumn{3}{c}{TMC13 v14~\cite{g_pcc}} & \multicolumn{3}{c}{TMC13 v23~\cite{g_pcc}} & \multicolumn{3}{c}{SparsePCAC~\cite{sparseW}} & \multicolumn{3}{c}{\textcolor{black}{TSC-PCAC~\cite{10693649}}} & \multicolumn{3}{c}{ScalablePCAC~\cite{10313579}} \\ \cline{3-17}           
&                                   & $128^2$  & $256^2$  & $512^2$ & $128^2$ & $256^2$ & $512^2$  & $128^2$ & $256^2$ & $512^2$  & \textcolor{black}{$128^2$} & \textcolor{black}{$256^2$}  & \textcolor{black}{$512^2$} & $128^2$ & $256^2$  & $512^2$ \\ \hline
\multirow{5}{*}{8iVFB} & longdress          & -21.15   & -18.94   & -18.41  &-16.06   &-13.42   &-12.42    & -11.78  & -11.84  & -11.54 &\textcolor{black}{\cmmnt{30.96}--}   &  \textcolor{black}{\cmmnt{28.32}--} &\textcolor{black}{\cmmnt{28.79}--}  &-0.94    &1.20      &1.01\\  
                       & redandblack        & -15.15   & -16.16   & -14.73  &-10.00   &-8.72    &-7.43     & -28.79  & -29.19  & -29.40 &\textcolor{black}{-18.01}   &  \textcolor{black}{-16.21} &\textcolor{black}{-16.72}  &0.30     &-0.22     &-0.91\\   
                       & loot               & -19.87   & -22.54   & -18.61  &-12.45   &-13.90   &-8.92     & -36.07  & -36.62  & -36.04 &\textcolor{black}{\cmmnt{6.16}--}   &  \textcolor{black}{\cmmnt{3.63}--} &\textcolor{black}{\cmmnt{4.72}--}  &-2.42    &-3.04     &-2.07 \\   
                       & soldier            & -35.09   & -35.89   & -33.43  &-28.45   &-27.89   &-25.47    & -28.83  & -27.81  & -27.76 &\textcolor{black}{-15.18}   &  \textcolor{black}{-13.68} &\textcolor{black}{-13.06}  &-11.45   &-10.92    &-10.58\\   \cline{2-17}
                       & \textbf{Average}            & \textbf{-22.81}   & \textbf{-23.38}   & \textbf{-21.23}  &\textbf{-16.74}   &\textbf{-16.00}   &\textbf{-13.60}    & \textbf{-26.36}  & \textbf{-26.37}  & \textbf{-26.19} &\textcolor{black}{\textbf{-16.60}}   &  \textcolor{black}{\textbf{-14.95}} &\textcolor{black}{\textbf{-14.89}}  &\textbf{-3.63}    &\textbf{-3.25}     &\textbf{-3.14}\\   \hline
\multirow{5}{*}{Owlii} & basketball  & -31.86   & -31.95   & -31.39  &-21.40   &-22.51   &-22.77    & -34.93  & -33.25  & -33.41 &\textcolor{black}{-7.81}   &  \textcolor{black}{-11.29} &\textcolor{black}{-9.90}  &-9.23    &-9.97     &-10.19\\    
                       & dancer             & -29.42   & -29.40   & -30.35  &-18.79   &-19.99   &-22.95    & -29.38  & -28.45  & -28.86 &\textcolor{black}{-7.28}   &  \textcolor{black}{-6.85} &\textcolor{black}{-8.31}  &-7.69    &-7.02     &-8.34\\    
                       & exercise           & -16.05   & -19.75   & -25.14  &-9.13    &-15.68   &-18.18    & -32.75  & -34.74  & -36.32 &\textcolor{black}{\cmmnt{12.41}--}   &  \textcolor{black}{\cmmnt{0.18}--} &\textcolor{black}{\cmmnt{-1.53}--}  &-3.88    &-4.75     &-7.66\\   
                       & model              & -23.65   & -25.93   & -29.18  &-19.20   &-19.02   &-21.22    & -16.01  & -19.71  & -23.24 &\textcolor{black}{\cmmnt{-3.05}--}   &  \textcolor{black}{\cmmnt{-2.70}--} &\textcolor{black}{\cmmnt{-2.45}--}  &-6.18    &-7.15     &-12.59\\   \cline{2-17}
                       & \textbf{Average}            & \textbf{-25.25}   & \textbf{-26.76}   & \textbf{-29.02}  &\textbf{-17.13}   &\textbf{-19.30}   &\textbf{-21.28}    & \textbf{-28.27}  & \textbf{-29.04}  & \textbf{-30.46} &\textcolor{black}{\textbf{-7.55}}   &  \textcolor{black}{\textbf{-9.07}} &\textcolor{black}{\textbf{-9.11}}  &\textbf{-6.75}    &\textbf{-7.22}     &\textbf{-9.70}\\  \hline
\multicolumn{17}{l}{{‘}--{’} indicates the test point cloud is selected into training dataset.}\\
\end{tabular}
\end{table*}
To accommodate the varying resolution requirements of display devices, we also conducted tests to evaluate the generalization of RO-PCAC across different resolutions. Experimental results of image resolutions of 128$\times$128, 256$\times$256 and 512$\times$512 are represented in Table.~\ref{resolutions}. We can see that the advantages of RO-PCAC method are still significant compared to other methods, which demonstrates that the neural network has learned suitable parameters to adapt to the rendering mechanism. \HXADD{We have also observed that although RO-PCAC outperforms all other baseline methods, the gains in BD-BR is reduced for lower-resolution multi-view images compared to 1024x1024 resolution. We attribute this reduction to the resolution discrepancy between training and testing images.}


\HXADD{In order to provide a more holistic evaluation of the proposed method, we implement an evaluation of using point-to-point Y PSNR as a 3D metric. As shown in Table~\ref{3dmetric}, compared with the results of image Y PSNR, the gains of all test point cloud decrease. For the 8iVFB dataset, RO-PCAC exhibits a performance that is slightly inferior to the state-of-the-art method for reconstructing the original point cloud (i.e., ScalablePCAC) by 1.17\%. These outcomes align with our research emphasis, as we aim to optimize for rendered image quality, which represents the ultimate output as perceived by the end-user in point cloud visualization scenarios. 
}

\begin{table}[htpb]
\centering
\caption{\HXADD{BD-BR (Y components) comparison with TMC13 v23, SparsePCAC and ScalablePCAC in point-to-point PSNR}}\label{3dmetric}
\setlength{\tabcolsep}{2.3pt} 
\color{black}
\begin{tabular}{clcccc}
\hline
\multirow{2}{*}{Class} & \multirow{2}{*}{Point Cloud} 
& \multicolumn{1}{c}{TMC13 v23} & \multicolumn{1}{c}{SparsePCAC} & \multicolumn{1}{c}{{TSC-PCAC}} & \multicolumn{1}{c}{ScalablePCAC } \\ \cline{3-6} 
                       &                              & Y            & Y           & Y             & Y      \\ \hline
\multirow{4}{*}{8iVFB} & longdress                    & -13.04      & -8.36        & \cmmnt{29.92}--& 9.10\\  
                       & redandblack                  & -16.12      & -18.58        & -13.99       & 1.46\\  
                       & loot                         & -3.94        & -33.76       & \cmmnt{8.63}--         & 2.33\\  
                       & soldier                      & -20.17        & -25.52        & -9.76         & -8.23\\ \hline
                       & \textbf{Average}             &\textbf{-13.32} &\textbf{-21.55}&\textbf{-11.88} &\textbf{1.17}\\ \hline
\multirow{4}{*}{Owlii} & basketball                   & -25.31        & -33.06        & -13.28         &-8.82\\  
                       & dancer                       & -26.14         & -26.76        & -9.71         & -7.05\\  
                       & exercise                     & -18.69        & -34.58       & \cmmnt{0.91}--        & -3.81\\ 
                       & model                        & -19.64         & -21.40       & \cmmnt{-6.87}--         & -9.02\\ \hline
                       & \textbf{Average}             &\textbf{-22.45}&\textbf{-28.95}&\textbf{-11.50} &\textbf{-7.18}\\ \hline
\multicolumn{6}{l}{{‘}--{’} indicates the test point cloud is selected into training dataset.}\\
\end{tabular}
\end{table}

\HXADD{Different distortion measures in the loss function can be used in our framework. Following the convention of image and video compression~\cite{zou2022devil,li2021deep, sheng2022temporal, cheng2020learned}, which uses Multi-Scale Structural Similarity (MS-SSIM) as loss function to enhance the subjective quality of images,
we also adopted the MS-SSIM as the loss function to train the model. Specifically, the distortion term is defined by $1-\text{MS} \text{-} \text{SSIM}(I,\hat{I})$. The results shown in Table~\ref{loss_msssim} indicate that when the model was optimized using MS-SSIM, there were increases in performance for all test point clouds compared to optimization with MSE. This indicates that the MS-SSIM optimized model improves the structural similarity of the rendered images of point clouds.}

\begin{table}[htpb]
\centering
\caption{\HXADD{BD-BR (MS-SSIM) comparison with TMC13 v23, SparsePCAC and ScalablePCAC using MS-SSIM optimized RO-PCAC}}\label{loss_msssim}
\setlength{\tabcolsep}{2.3pt} 
\color{black}
\begin{tabular}{clcccc}
\hline
\multirow{2}{*}{Class} & \multirow{2}{*}{Point Cloud} 
& \multicolumn{1}{c}{TMC13 v23} & \multicolumn{1}{c}{SparsePCAC} & \multicolumn{1}{c}{{TSC-PCAC}} & \multicolumn{1}{c}{ScalablePCAC } \\ \cline{3-6} 
                       &                              & MS-SSIM            & MS-SSIM            & MS-SSIM              & MS-SSIM       \\ \hline
\multirow{4}{*}{8iVFB} & longdress                    & -11.36             & -16.21             & \cmmnt{29.92}--      & -10.40\\  
                       & redandblack                  & 10.51              & -42.18             & -39.29               & -13.96\\  
                       & loot                         & -13.52             & -43.12             & \cmmnt{8.63}--       & -9.31\\  
                       & soldier                      & -38.17             & -31.17             & -1.76                & -18.42\\ \hline
                       & \textbf{Average}             &\textbf{-13.14}     &\textbf{-33.17}     &\textbf{-20.53}       &\textbf{-13.02}\\ \hline
\multirow{4}{*}{Owlii} & basketball                   & -36.21             & -34.01             & -3.28                & -24.12\\  
                       & dancer                       & -33.11             & -36.58             & -5.13                & -17.09\\  
                       & exercise                     & -29.14             & -34.29             & \cmmnt{0.91}--       & -23.31\\ 
                       & model                        & -36.64             & -21.12             & \cmmnt{-6.87}--      & -20.62\\ \hline
                       & \textbf{Average}             &\textbf{-33.78}&\textbf{-31.50}&\textbf{-4.21} &\textbf{-21.29}\\ \hline
\multicolumn{6}{l}{{‘}--{’} indicates the test point cloud is selected into training dataset.}\\
\end{tabular}
\end{table}

\HXADD{Except for the human-based models, RO-PCAC can also compress indoor scene point cloud such as ScanNet~\cite{dai2017scannet}. This experiment reports the compression performance over G-PCC TMC13 v23. To avoid presenting a simplistic average performance of the 100 test point clouds, we report the average performance at increments of 25, specifically for the first 25, 50, 75, and all 100 point clouds, to demonstrate the robustness of our method across various indoor scenes.
The results presented in Table~\ref{scans_test} demonstrate that RO-PCAC outperforms G-PCC TMC13 v23 across all frame sets.}

\begin{table}[htpb]
\centering
\setlength{\tabcolsep}{10pt} 
\color{black}
\caption{\HXADD{BD-BR (Y component, YUV component and MS-SSIM) comparison with TMC13 v23 of ScanNet under different frame sets}}
\label{scans_test}
\begin{tabular}{ccccc}
\hline
\multirow{2}{*}{Metrics} & \multicolumn{4}{c}{Frames} \\ \cline{2-5} 
                              & 25   & 50   & 75   & 100   \\ \hline
Y                           & -10.48     &   -9.60   & -7.87  &   -8.61    \\ \hline
YUV                           &  -7.03    &    -6.03  & -4.12 &    -5.19   \\ \hline
MS-SSIM                          &  -10.45    &  -11.36    &  -9.12    &   -10.19    \\ \hline
\end{tabular}
\end{table}

\subsubsection{Subjective Comparison.} 
\begin{figure*}[htbp]
	    \centering
            \subcaptionbox{\parbox{3cm}{\vspace{-7pt}\centering Ground Truth \\ \textit{\textbf{Dancer}}}}{
            \includegraphics[width=29mm]
	    {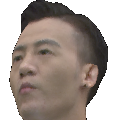}
            }
            \subcaptionbox{\parbox{3cm}{\vspace{-7pt}\centering TMC13 v23 \\ Y-PSNR: 36.43 dB \\ YUV-PSNR: 37.55 dB \\ MS-SSIM: 0.9973 \\ bitrate: 0.1474 bpp}}{
            \includegraphics[width=29mm]
	    {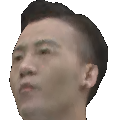}
            }
            \subcaptionbox{\parbox{3cm}{\vspace{-7pt}\centering SparsePCAC \\ Y-PSNR: 34.82 dB \\ YUV-PSNR: 35.90 dB \\ MS-SSIM: 0.9960 \\ bitrate: 0.0999 bpp}}{
            \includegraphics[width=29mm]
	    {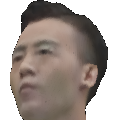}
            }
            \subcaptionbox{\parbox{3cm}{\vspace{-7pt}\centering ScalablePCAC \\ Y-PSNR: 37.32 dB \\ YUV-PSNR: 38.38 dB \\ MS-SSIM: 0.9978 \\ bitrate: 0.1472 bpp}}{
            \includegraphics[width=29mm]
	    {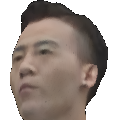}
            }
            \subcaptionbox{\parbox{3cm}{\vspace{-7pt}\centering RO-PCAC \\ Y-PSNR: 37.76 dB \\ YUV-PSNR: 38.80 dB \\ MS-SSIM: 0.9980 \\ bitrate: 0.1523 bpp}}{
            \includegraphics[width=29mm]
	    {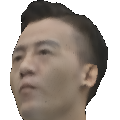}
            }
            \vspace{10pt}

            \subcaptionbox{\parbox{3cm}{\vspace{-7pt}\centering Ground Truth \\ \textit{\textbf{Model}}}}{
                \includegraphics[width=29mm]
    	    {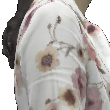}
                }
            \subcaptionbox{\parbox{3cm}{\vspace{-7pt}\centering TMC13 v23 \\ Y-PSNR: 35.50 dB \\ YUV-PSNR: 36.57 dB \\ MS-SSIM: 0.9980 \\ bitrate: 0.2251 bpp}}{
            \includegraphics[width=29mm]
	    {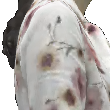}
            }
            \subcaptionbox{\parbox{3cm}{\vspace{-7pt}\centering SparsePCAC \\ Y-PSNR: 34.99 dB \\ YUV-PSNR: 36.05 dB \\ MS-SSIM: 0.9979 \\ bitrate: 0.2114 bpp}}{
            \includegraphics[width=29mm]
	    {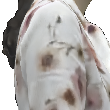}
            }
            \subcaptionbox{\parbox{3cm}{\vspace{-7pt}\centering ScalablePCAC \\ Y-PSNR: 35.95 dB \\ YUV-PSNR: 37.02 dB \\ MS-SSIM: 0.9984 \\ bitrate: 0.2201 bpp}}{
            \includegraphics[width=29mm]
	    {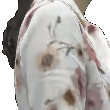}
            }
            \subcaptionbox{\parbox{3cm}{\vspace{-7pt}\centering RO-PCAC \\ Y-PSNR: 36.43 dB \\ YUV-PSNR: 37.49 dB \\ MS-SSIM: 0.9986 \\ bitrate: 0.2106 bpp}}{
            \includegraphics[width=29mm]
	    {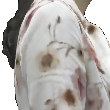}
            }

            \vspace{10pt}
            \subcaptionbox{\parbox{3cm}{\vspace{-7pt}\centering Ground Truth \\ \textit{\textbf{RedandBlack}}}}{
            \includegraphics[width=29mm]
	    {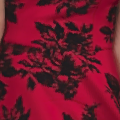}
            }
            \subcaptionbox{\parbox{3cm}{\vspace{-7pt}\centering TMC13 v23 \\ Y-PSNR: 33.41 dB \\ YUV-PSNR: 33.97 dB \\ MS-SSIM: 0.9943 \\ bitrate: 0.1039 bpp}}{
            \includegraphics[width=29mm]
	    {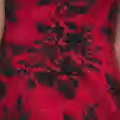}
            }
            \subcaptionbox{\parbox{3cm}{\vspace{-7pt}\centering SparsePCAC \\ Y-PSNR: 32.97 dB \\ YUV-PSNR: 32.60 dB \\ MS-SSIM: 0.9903 \\ bitrate: 0.0835 bpp}}{
            \includegraphics[width=29mm]
	    {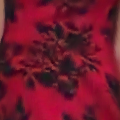}
            }
            \subcaptionbox{\parbox{3cm}{\vspace{-7pt}\centering ScalablePCAC \\ Y-PSNR: 34.63 dB \\ YUV-PSNR: 34.47 dB \\ MS-SSIM: 0.9927 \\ bitrate: 0.1107 bpp}}{
            \includegraphics[width=29mm]
	    {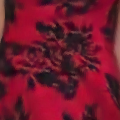}
            }
            \subcaptionbox{\parbox{3cm}{\vspace{-7pt}\centering RO-PCAC \\ Y-PSNR: 34.75 dB \\ YUV-PSNR: 34.57 dB \\ MS-SSIM: 0.9936 \\ bitrate: 0.1008 bpp}}{
            \includegraphics[width=29mm]
	    {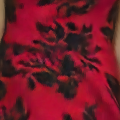}
            }

            \caption{Visual quality comparison of reconstructed point clouds among TMC13v23, SparsePCAC, ScalablePCAC and RO-PCAC. Point clouds \textit{Dancer}, \textit{Model}, and \textit{RedandBlack} are arranged from top to bottom.}\label{subjective}
\end{figure*}

The rendered images of reconstructed point clouds for G-PCC, SparsePCAC, ScalablePCAC and RO-PCAC are presented in Fig.~\ref{subjective}. To better illustrate the visual contrast, we enlarged portions of the rendered images. Furthermore, to ensure a balanced evaluation, similar rate points for each method were selected for comparison.

It is evident that the rendered images from G-PCC TMC13v23 exhibit noticeable blocking artifacts, particularly in the eyes of \textit{Dancer}, flowers on the arm of \textit{Model} and the black pattern of \textit{RedandBlack}. SparsePCAC, while effective in reducing these artifacts, introduces a new issue: it over-smooths the texture details, leading to a loss of texture details. In contrast, our proposed method offers visually appealing reconstructions that retain rich texture details. The blocking artifacts are significantly reduced, and the native textures are preserved effectively. This balance between artifact reduction and texture preservation demonstrates the method's efficacy in point cloud attribute compression, providing superior visual quality compared to the other methods evaluated.

\subsection{Ablation Study}
\begin{figure}[htpb]
	    \centering
            \subcaptionbox{8iVFB}{
            \includegraphics[width=41mm]
	    {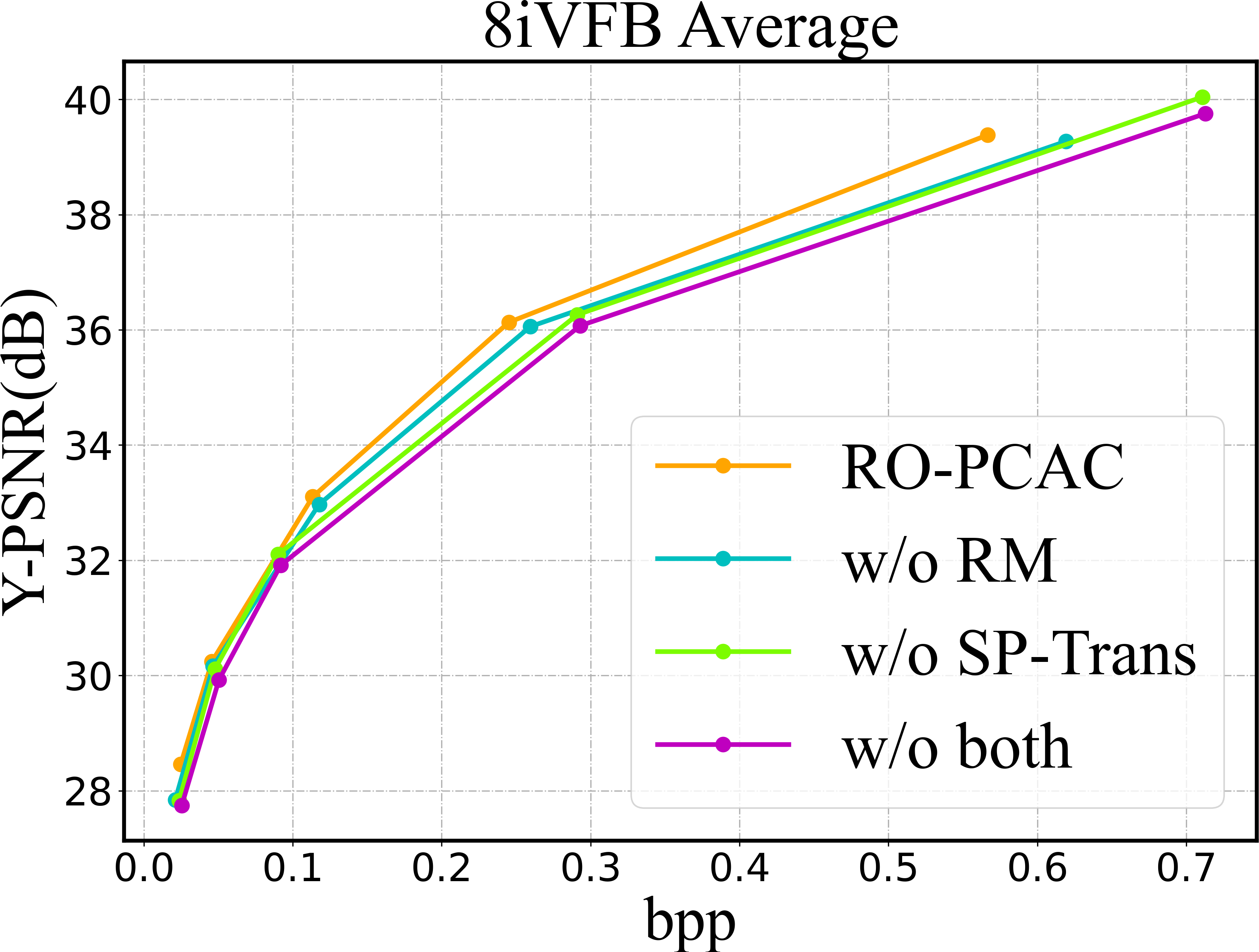}
            }
            \subcaptionbox{Owlii}{
            \includegraphics[width=41mm]
	    {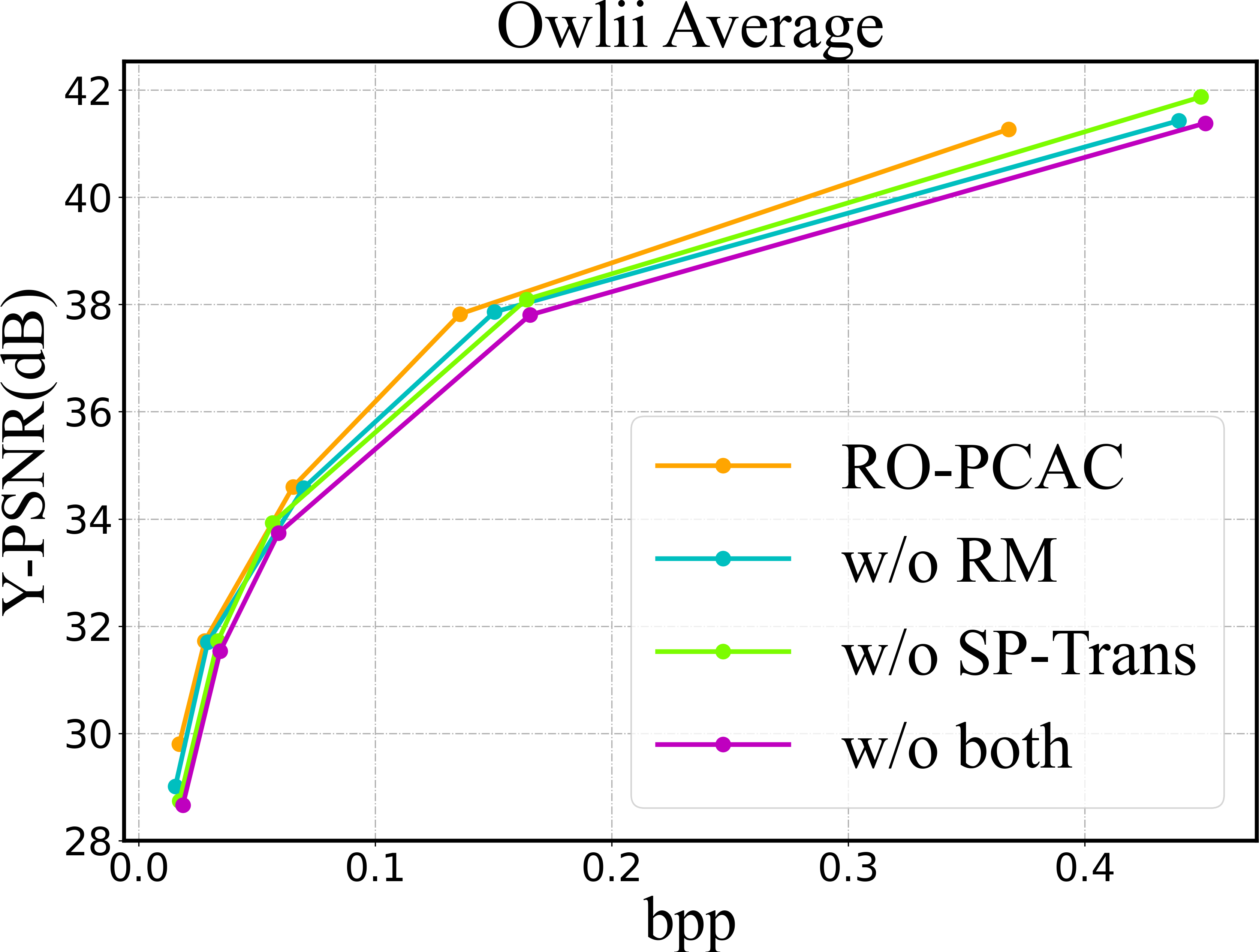}
            }
            \caption{R-D curves of different ablation configurations.}\label{ab}
\end{figure}

\HXADD{\subsubsection{Ablation Study of SP-Trans and Rendering Module}}
In this section, we present an ablation study to evaluate the contributions of SP-Trans and Rendering Module in our proposed model. 
The ablation study is conducted by \HXDEL{incrementally removing} \HXADD{replacing or removing} components from the full model and measuring the resulting performance degradation.
The outcomes of the ablation study are shown in Fig.~\ref{ab} and Table.~\ref{abla}, which provide a comparative analysis of the model's performance \HXDEL{without SP-Trans and Rendering Module (i.e., w/o SP-Trans and w/o RM). }
\HXADD{through replacing SP-Trans with SConv and removing Rendering Module (i.e., w/o SP-Trans and w/o RM).} Specially, when the Rendering Module is removed, point-to-point loss is used to train the corresponding model.
The exclusion of SP-Trans, Rendering Module and both of them resulted in a decrement in BD-BR by \HXDEL{9.86\%}\HXADD{6.66\%}, 7.38\% and \HXDEL{17.02\%}\HXADD{13.96\%} in Y component in 8iVFB, respectively. 
Similarly, on the Owlii dataset, the absence of these components led to decrements of \HXDEL{10.94\%}\HXADD{8.32\%}, 8.83\%, and \HXDEL{20.20\%} \HXADD{16.94\%} in the BD-BR for the Y component, respectively, which demonstrates the effectiveness of SP-Trans and the Rendering Module.
\begin{table}[htpb]
\centering
\caption{Ablation Studies on Modular Components in RO-PCAC}\label{abla}
\begin{tabular}{clccc}
\hline
\multirow{2}{*}{Class} & \multirow{2}{*}{Point Cloud} & \multicolumn{1}{c}{w/o SP-Trans} & \multicolumn{1}{c}{w/o RM} & \multicolumn{1}{c}{w/o both} \\ \cline{3-5} 
                       &                              & Y           & Y          & Y                 \\ \hline
\multirow{4}{*}{8iVFB} & longdress                    & \color{black}1.59        & 2.21       & \color{black}3.54\\  
                       & redandblack                  & \color{black}3.71        & 6.38       & \color{black}10.20\\  
                       & loot                         & \color{black}8.18       & 11.22      & \color{black}20.24\\  
                       & soldier                      & \color{black}13.23       & 9.65       & \color{black}21.92\\ \hline
                       & \textbf{Average}             &\color{black}\textbf{6.66}&\textbf{7.38}&\color{black}\textbf{13.96}\\ \hline
\multirow{4}{*}{Owlii} & basketball                   &\color{black} 10.11       & 9.27      & \color{black}19.14\\  
                       & dancer                       &\color{black} 6.15        & 8.07      & \color{black}14.66\\  
                       & exercise                     &\color{black} 13.79       & 11.28      & \color{black}24.19\\ 
                       & model                        &\color{black} 3.21        & 6.69       & \color{black}9.76\\ \hline
                       & \textbf{Average}             &\color{black}\textbf{8.32}&\textbf{8.83}&\color{black}\textbf{16.94}\\ \hline
\end{tabular}
\end{table}

\HXADD{\subsubsection{Ablation Study of Variations in SP-Trans}
In this section, we present an ablation study to evaluate the contributions of variations in SP-Trans, compared with the typical point transformer~\cite{Zhao_2021_ICCV}. The ablation study is conducted by individually replacing the variations in SP-Trans with the corresponding components in~\cite{Zhao_2021_ICCV} and measuring the resulting performance degradation. 
The outcomes of the ablation study are shown in Table.~\ref{vars}, which provide a comparative analysis of the model's performance through replacing cosine with softmax similarity measurement and replacing neighbor searching in 3D window with K nearest neighbors (K=16 as suggested by~\cite{Zhao_2021_ICCV}) (i.e., w/ softmax and w/ KNN). 
Specially, in order to demonstrate the efficient neighbor searching in 3D window of SP-Trans, we test the added runtime by using KNN.
The replacement with softmax and KNN resulted in a decrement in BD-BR by 5.63\% and 3.65\% in Y component in 8iVFB, respectively. Similarly, on the Owlii dataset, the absence of these components led to decrements of 4.80\% and 3.78\% in the BD-BR for the Y component, respectively. In addition, the total runtime of encoder and decoder by using KNN added 0.41s (43\%) and 0.31s (33\%), respectively.}
\begin{table}[htpb]
\centering
\caption{\HXADD{Ablation Studies on Variations in SP-Trans}}\label{vars}
\color{black}
\begin{tabular}{clc|cc}
\hline
\multirow{2}{*}{ Class}& \multirow{2}{*}{ Point Cloud} & \multicolumn{1}{c}{w/ softmax} & \multicolumn{2}{c}{w/ KNN}  \\ \cline{3-5} 
                       &                              & Y           & Y           & Added Runtime                   \\ \hline
\multirow{4}{*}{8iVFB} & longdress                    & 3.44        & 4.34       & 0.36s (38\%)\\  
                       & redandblack                  & 5.26        & 6.67       & 0.58s (43\%)\\  
                       & loot                         & 4.26        & 3.21       & 0.27s (45\%)\\  
                       & soldier                      & 5.17        & 3.65        & 0.43s (47\%)\\ \hline
                       & \textbf{Average}             &\textbf{5.63}&\textbf{4.47}&\textbf{0.41s} (43\%)\\ \hline
\multirow{4}{*}{Owlii} & basketball                   & 2.46       & 3.26        & 0.39s (41\%)\\  
                       & dancer                       & 5.31        & 3.37        & 0.28s (22\%)\\  
                       & exercise                     & 5.16      & 6.28       & 0.21s (33\%)\\ 
                       & model                        & 6.26        & 2.21       & 0.37s (46\%)\\ \hline
                       & \textbf{Average}             &\textbf{4.80}&\textbf{3.78}&\textbf{0.31 (33\%)}\\ \hline
    \multicolumn{5}{l}{The relative added percentage is indicated in parentheses.}\\
\end{tabular}
\end{table}

\HXADD{\subsubsection{Ablation Study of Number of Projections}
In this section, we present an ablation study to evaluate the influence of the number of projection images on the performance. This study involves varying the number of projections used during the training and testing phases and presenting corresponding compression performances in dataset 8iVFB. For training projection, the numbers are set to 2, 4, 6, 8 and 14 respectively. For the 2 projections, the elevations is $0^\circ$ and azimuths are $0^\circ$ and $180^\circ$. For 4, 6, 8 and 14 projections, 2 views are with the elevations of $90^\circ$ and $270^\circ$ and azimuths of $0^\circ$, and the other views are with the elevations of $0^\circ$ and azimuths ranging from  $0^\circ$ to $360^\circ$ with intervals of $180^\circ$, $90^\circ$, $60^\circ$ and $30^\circ$, respectively.
For testing projection, the numbers are set to 2, 4, 6 and 12 respectively.  In order to avoid overlapping perspectives adopted in training, the elevations is $0^\circ$ and azimuths ranges from  $15^\circ$ to $360^\circ$ with intervals of $180^\circ$, $90^\circ$, $60^\circ$ and $30^\circ$, respectively.
The outcomes of the ablation study are shown in Table.~\ref{numofproj}, it is evident that as the number of training projections increases, the BD-BR values become more negative. However, when the number of training projections is greater than 4, the gains tend to decrease. This can be observed in the table where the BD-BR values approach -17.20\% for training projections of 6, 8, and 12, with only marginal differences between these values. This suggests that beyond a certain point, which in this case is around 8 training projections, the additional training data does not significantly contribute to further performance improvements. This trend is consistent across all testing projection counts. Moreover, for a well-trained model, as the number of testing projections increases from 2 to 12, the performance remains stable with only slight changes, which indicates that the model is robust to testing projection.}

\HXADD{The plateauing and stabilizing of gains can be attributed to the strong supervision in training. With the convention of the geometry losslessly encoded, the geometry of the decoded point clouds are the same as those of original point clouds. Furthermore, the pixels between the images respectively rendered from the original and decoded point cloud have accurate one-to-one correspondences. Based on these strong supervisions, the
gradients calculated from MSE of images are allowed to be propagated from pixel values backward to latent features of exact points, making the network is easy to converge and extract effective features. Therefore, the number of projection images for training does not need too many, as long as the images can basically cover the point cloud. If the number of projection images for training is too small to cover the point cloud, features of partial points will not be supervised, and thus the trained network generates quality relatively lowered images. If more projection images are used, there also is no significant improvement in performance. }

\begin{table}[htpb]
\centering
\setlength{\tabcolsep}{7pt} 
\caption{\HXADD{BD-BR (Y component) comparison with TMC13 v23 of 8iVFB under different numbers of projections}\label{numofproj}}
\color{black}
\begin{tabular}{@{}cccccc@{}}
\toprule
{ }   & \multicolumn{4}{c}{{ Training projections}}                                       \\ \cmidrule(l){2-6} 
\multirow{-2.5}{*}{{ \begin{tabular}[c]{@{}c@{}}Testing \\ projections\end{tabular}}} &
  { 2} &
  { 4} &
  { 6} &
  { 8} &
  { 14} \\ \midrule
{ 2}  & {  -8.28} & { -13.19} & {  -15.90}&{ -17.10}  & { -17.20} \\ \midrule
{ 4}  & {  -7.78} & { -13.13} & { -15.92}&{ -17.01}  & { -17.14} \\ \midrule
{ 6}  & {  -8.13} & { -12.59} & { -15.77}&{ -17.13}  & { -17.14} \\ \midrule
{ 12} & { -7.97} & {  -13.01} & { -15.89}&{ -17.11}  & { -17.17} \\ \bottomrule
\end{tabular}
\end{table}

\subsection{Compression Complexity}
In this section, we evaluated the time complexity of encoding and decoding for various codecs. Each point cloud is processed sequentially on a fixed GPU and CPU, with the operating system being Ubuntu 20.04. The encoding and decoding times are the average times for all point clouds in the testing dataset.

Table~\ref{time} illustrates the detailed time consumption in the encoder and decoder. For learning-based methods, {‘}Y and Z analysis{’} represents the time of generating the latent and hyperprior features in the encoder and {‘}Y and Z synthesis{’} represents the time of reconstructing point clouds through latent and hyperprior features in the decoder. {‘}Y and Z compressing{’} and {‘}Y and Z decompressing{’} represent the time of compressing and decompressing the latent and hyperprior features by the arithmetic encoder and arithmetic decoder, respectively. {‘}B.{’} and {‘}E.{’} represent the base layer and enhancement layer in ScalablePCAC, respectively. We can observe that RO-PCAC has a comparable encoding/decoding time consumption with G-PCC, which are 4.42s/4.79s and 3.92s/3.68s, respectively. However, the encoding/decoding times of SparsePCAC and ScalablePCAC are much longer than RO-PCAC, which are 20.79s/201.52s and 53.68s/351.54s, respectively. This is due to the entropy model they adopted needing to take previous decoded latent features as input to generate the entropy model parameters of the current features to be encoded. In contrast, RO-PCAC directly outputs the entropy parameters of all latent features and encode them simultaneously, having a huge time saving. In addition, in the learning included methods, SparsePCAC has the shortest time of feature analysis and synthesis, which are 0.17s and 0.44s, respectively. Compared with SparsePCAC, RO-PCAC only has time addition of 0.1s and 0.09s, respectively. This is attributed to the efficient neighbor searching in SP-Trans, which has the same time complexity as that in sparse convolution.

\begin{table}[htpb]
\centering
\caption{Runtime (seconds) analysis}\label{time}
\begin{tabular}{cccc}
\hline
\multicolumn{2}{c}{Method}                       & Encoder & Decoder \\ \hline
\multicolumn{2}{c}{G-PCC TMC13v23}               & 3.92        & 3.68       \\ \hline
\multirow{5}{*}{SparsePCAC}      & Y and Z analysis       & 0.17        &  -        \\ \cline{2-4} 
                                 & Y and Z synthesis        & -        & 0.44        \\ \cline{2-4} 
                                 & Y and Z compressing        & 20.62        & -        \\ \cline{2-4} 
                                 & Y and Z decompressing        & -        & 201.08        \\ \cline{2-4} 
                                 & total & 20.79        & 201.52       \\ \hline

\multirow{5}{*}{\color{black}TSC-PCAC}        & \color{black} Y and Z analysis       & \color{black}0.57        & \color{black} -        \\ \cline{2-4} 
                                 & \color{black}Y and Z synthesis        &\color{black} -        &\color{black} 0.84        \\ \cline{2-4} 
                                 & \color{black}Y and Z compressing        &\color{black} 5.62        &\color{black} -        \\ \cline{2-4} 
                                 & \color{black}Y and Z decompressing        & \color{black}-        & \color{black}6.08        \\ \cline{2-4} 
                                 & \color{black}total & \color{black}5.79        & \color{black}6.52       \\ \hline

\multirow{6}{*}{ScalablePCAC}    & B. G-PCC       & 0.85        &  0.41        \\ \cline{2-4}
                                 & E. Y and Z analysis       & 5.77        &  -        \\ \cline{2-4} 
                                 & E. Y and Z synthesis        & -        & 4.08        \\ \cline{2-4} 
                                 & E. Y and Z compressing        & 47.06        & -        \\ \cline{2-4} 
                                 & E. Y and Z decompressing        & -        & 347.05        \\ \cline{2-4} 
                                 & total & 53.68        & 351.54        \\ \hline
\multirow{5}{*}{RO-PCAC} & Y and Z analysis       & 0.27        &  -        \\ \cline{2-4} 
                                 & Y and Z synthesis        & -        & 0.53        \\ \cline{2-4} 
                                 & Y and Z compressing        & 4.15        & -        \\ \cline{2-4} 
                                 & Y and Z decompressing        & -        & 4.56        \\ \cline{2-4} 
                                 & total & 4.42        & 4.79        \\ \hline
\end{tabular}
\end{table}

\section{conclusion}
\label{sec:con}
We have presented RO-PCAC, a deep learning framework for rendering-oriented point cloud compression. By integrating the differentiable rendering into the compression process, RO-PCAC balances compression efficiency with the quality of rendered images. The introduction of SP-Trans, leveraging sparse tensor representations, efficiently enhances feature analysis and synthesis by capturing the complex interdependencies in regions of varying density. Our experimental results indicate that RO-PCAC achieves the state-of-the-art performance, demonstrating improvements in compression efficiency compared to the MPEG G-PCC standard and other contemporary learning-based and hybrid methods. \HXADD{However, it is noted that the performance gain over ScalablePCAC is relatively small, less than 10\%. This motivates us to further explore more strategies to enhance RO-PCAC to achieve more gains in the future. Incorporating resolution as a variable in the framework to enhance performance for low-resolution images is also a valuable direction for future work, which would be beneficial for bandwidth limited scenarios where transmitting low-resolution images may be necessary. Moreover, the fixed weights in alpha-blending (i.e., Eq.~\eqref{color}) in the rendering module may limit the model's flexibility. Investigating adaptive weight strategies could be a valuable research to improve the rendering pipeline. This could include learning the weights dynamically based on the data or integrating attention mechanisms to allocate weights in a manner that better captures the salient features of the point cloud data.}


\bibliographystyle{IEEEtran}
\bibliography{IEEEabrv,./ref.bib}

\begin{thebibliography}{10}
\providecommand{\url}[1]{#1}
\csname url@samestyle\endcsname
\providecommand{\newblock}{\relax}
\providecommand{\bibinfo}[2]{#2}
\providecommand{\BIBentrySTDinterwordspacing}{\spaceskip=0pt\relax}
\providecommand{\BIBentryALTinterwordstretchfactor}{4}
\providecommand{\BIBentryALTinterwordspacing}{\spaceskip=\fontdimen2\font plus
\BIBentryALTinterwordstretchfactor\fontdimen3\font minus \fontdimen4\font\relax}
\providecommand{\BIBforeignlanguage}[2]{{%
\expandafter\ifx\csname l@#1\endcsname\relax
\typeout{** WARNING: IEEEtran.bst: No hyphenation pattern has been}%
\typeout{** loaded for the language `#1'. Using the pattern for}%
\typeout{** the default language instead.}%
\else
\language=\csname l@#1\endcsname
\fi
#2}}
\providecommand{\BIBdecl}{\relax}
\BIBdecl

\bibitem{7434610}
R.~Mekuria, K.~Blom, and P.~Cesar, ``Design, implementation, and evaluation of a point cloud codec for tele-immersive video,'' \emph{IEEE Transactions on Circuits and Systems for Video Technology}, vol.~27, no.~4, pp. 828--842, 2017.

\bibitem{10.1145/3145690.3145729}
\BIBentryALTinterwordspacing
J.~D. Stets, Y.~Sun, W.~Corning, and S.~W. Greenwald, ``Visualization and labeling of point clouds in virtual reality,'' in \emph{SIGGRAPH Asia 2017 Posters}, ser. SA '17.\hskip 1em plus 0.5em minus 0.4em\relax New York, NY, USA: Association for Computing Machinery, 2017. [Online]. Available: \url{https://doi.org/10.1145/3145690.3145729}
\BIBentrySTDinterwordspacing

\bibitem{7123038}
F.~Lozes, A.~Elmoataz, and O.~Lezoray, ``Pde-based graph signal processing for 3-d color point clouds: Opportunities for cultural heritage,'' \emph{IEEE Signal Processing Magazine}, vol.~32, no.~4, pp. 103--111, 2015.

\bibitem{9503405}
W.~Zhu, Y.~Xu, D.~Ding, Z.~Ma, and M.~Nilsson, ``Lossy point cloud geometry compression via region-wise processing,'' \emph{IEEE Transactions on Circuits and Systems for Video Technology}, vol.~31, no.~12, pp. 4575--4589, 2021.

\bibitem{9500191}
B.~Zhao, W.~Lin, and C.~Lv, ``Fine-grained patch segmentation and rasterization for 3-d point cloud attribute compression,'' \emph{IEEE Transactions on Circuits and Systems for Video Technology}, vol.~31, no.~12, pp. 4590--4602, 2021.

\bibitem{10024999}
D.~T. Nguyen and A.~Kaup, ``Lossless point cloud geometry and attribute compression using a learned conditional probability model,'' \emph{IEEE Transactions on Circuits and Systems for Video Technology}, vol.~33, no.~8, pp. 4337--4348, 2023.

\bibitem{9957096}
P.~Gao, S.~Luo, and M.~Paul, ``Rate-distortion modeling for bit rate constrained point cloud compression,'' \emph{IEEE Transactions on Circuits and Systems for Video Technology}, vol.~33, no.~5, pp. 2424--2438, 2023.

\bibitem{8816692}
S.~Gu, J.~Hou, H.~Zeng, H.~Yuan, and K.-K. Ma, ``3d point cloud attribute compression using geometry-guided sparse representation,'' \emph{IEEE Transactions on Image Processing}, vol.~29, pp. 796--808, 2020.

\bibitem{8949735}
S.~Gu, J.~Hou, H.~Zeng, and H.~Yuan, ``3d point cloud attribute compression via graph prediction,'' \emph{IEEE Signal Processing Letters}, vol.~27, pp. 176--180, 2020.

\bibitem{8571288}
S.~Schwarz, M.~Preda, V.~Baroncini, M.~Budagavi, P.~Cesar, P.~A. Chou, R.~A. Cohen, M.~Krivokuća, S.~Lasserre, Z.~Li, J.~Llach, K.~Mammou, R.~Mekuria, O.~Nakagami, E.~Siahaan, A.~Tabatabai, A.~M. Tourapis, and V.~Zakharchenko, ``Emerging mpeg standards for point cloud compression,'' \emph{IEEE Journal on Emerging and Selected Topics in Circuits and Systems}, vol.~9, no.~1, pp. 133--148, 2019.

\bibitem{8700674}
E.~S. Jang, M.~Preda, K.~Mammou, A.~M. Tourapis, J.~Kim, D.~B. Graziosi, S.~Rhyu, and M.~Budagavi, ``Video-based point-cloud-compression standard in mpeg: From evidence collection to committee draft [standards in a nutshell],'' \emph{IEEE Signal Processing Magazine}, vol.~36, no.~3, pp. 118--123, 2019.

\bibitem{8945224}
H.~Liu, H.~Yuan, Q.~Liu, J.~Hou, and J.~Liu, ``A comprehensive study and comparison of core technologies for mpeg 3-d point cloud compression,'' \emph{IEEE Transactions on Broadcasting}, vol.~66, no.~3, pp. 701--717, 2020.

\bibitem{9418789}
J.~Wang, D.~Ding, Z.~Li, and Z.~Ma, ``Multiscale point cloud geometry compression,'' in \emph{2021 Data Compression Conference (DCC)}, 2021, pp. 73--82.

\bibitem{wang2022sparse}
J.~Wang, D.~Ding, Z.~Li, X.~Feng, C.~Cao, and Z.~Ma, ``Sparse tensor-based multiscale representation for point cloud geometry compression,'' \emph{IEEE Transactions on Pattern Analysis and Machine Intelligence}, vol.~45, no.~7, pp. 9055--9071, 2022.

\bibitem{huang2020octsqueeze}
L.~Huang, S.~Wang, K.~Wong, J.~Liu, and R.~Urtasun, ``Octsqueeze: Octree-structured entropy model for lidar compression,'' in \emph{Proceedings of the IEEE/CVF conference on computer vision and pattern recognition}, 2020, pp. 1313--1323.

\bibitem{que2021voxelcontext}
Z.~Que, G.~Lu, and D.~Xu, ``Voxelcontext-net: An octree based framework for point cloud compression,'' in \emph{Proceedings of the IEEE/CVF Conference on Computer Vision and Pattern Recognition}, 2021, pp. 6042--6051.

\bibitem{fu2022octattention}
C.~Fu, G.~Li, R.~Song, W.~Gao, and S.~Liu, ``Octattention: Octree-based large-scale contexts model for point cloud compression,'' in \emph{Proceedings of the AAAI conference on artificial intelligence}, vol.~36, no.~1, 2022, pp. 625--633.

\bibitem{song2023efficient}
R.~Song, C.~Fu, S.~Liu, and G.~Li, ``Efficient hierarchical entropy model for learned point cloud compression,'' in \emph{Proceedings of the IEEE/CVF Conference on Computer Vision and Pattern Recognition}, 2023, pp. 14\,368--14\,377.

\bibitem{deeppcac}
X.~Sheng, L.~Li, D.~Liu, Z.~Xiong, Z.~Li, and F.~Wu, ``Deep-pcac: An end-to-end deep lossy compression framework for point cloud attributes,'' \emph{IEEE Transactions on Multimedia}, vol.~24, pp. 2617--2632, 2022.

\bibitem{9191180}
M.~Quach, G.~Valenzise, and F.~Dufaux, ``Folding-based compression of point cloud attributes,'' in \emph{2020 IEEE International Conference on Image Processing (ICIP)}, 2020, pp. 3309--3313.

\bibitem{sparseW}
J.~Wang and Z.~Ma, ``Sparse tensor-based point cloud attribute compression,'' in \emph{2022 IEEE 5th International Conference on Multimedia Information Processing and Retrieval (MIPR)}, 2022, pp. 59--64.

\bibitem{10313579}
J.~Zhang, J.~Wang, D.~Ding, and Z.~Ma, ``Scalable point cloud attribute compression,'' \emph{IEEE Transactions on Multimedia}, pp. 1--11, 2023.

\bibitem{10474112}
V.~F. Figueiredo, R.~L. de~Queiroz, P.~A. Chou, and L.~S. Lopes, ``Embedded coding of point cloud attributes,'' \emph{IEEE Signal Processing Letters}, vol.~31, pp. 890--893, 2024.

\bibitem{9257015}
A.~Javaheri, C.~Brites, F.~Pereira, and J.~Ascenso, ``Point cloud rendering after coding: Impacts on subjective and objective quality,'' \emph{IEEE Transactions on Multimedia}, vol.~23, pp. 4049--4064, 2021.

\bibitem{Zhao_2021_ICCV}
H.~Zhao, L.~Jiang, J.~Jia, P.~H. Torr, and V.~Koltun, ``Point transformer,'' in \emph{Proceedings of the IEEE/CVF International Conference on Computer Vision (ICCV)}, October 2021, pp. 16\,259--16\,268.

\bibitem{NEURIPS2022_d78ece66}
\BIBentryALTinterwordspacing
X.~Wu, Y.~Lao, L.~Jiang, X.~Liu, and H.~Zhao, ``Point transformer v2: Grouped vector attention and partition-based pooling,'' in \emph{Advances in Neural Information Processing Systems}, S.~Koyejo, S.~Mohamed, A.~Agarwal, D.~Belgrave, K.~Cho, and A.~Oh, Eds., vol.~35.\hskip 1em plus 0.5em minus 0.4em\relax Curran Associates, Inc., 2022, pp. 33\,330--33\,342. [Online]. Available: \url{https://proceedings.neurips.cc/paper_files/paper/2022/file/d78ece6613953f46501b958b7bb4582f-Paper-Conference.pdf}
\BIBentrySTDinterwordspacing

\bibitem{guo2021pct}
M.-H. Guo, J.-X. Cai, Z.-N. Liu, T.-J. Mu, R.~R. Martin, and S.-M. Hu, ``Pct: Point cloud transformer,'' \emph{Computational Visual Media}, vol.~7, pp. 187--199, 2021.

\bibitem{choy20194d}
C.~Choy, J.~Gwak, and S.~Savarese, ``4d spatio-temporal convnets: Minkowski convolutional neural networks,'' in \emph{Proceedings of the IEEE Conference on Computer Vision and Pattern Recognition}, 2019, pp. 3075--3084.

\bibitem{gwak2020gsdn}
J.~Gwak, C.~B. Choy, and S.~Savarese, ``Generative sparse detection networks for 3d single-shot object detection,'' in \emph{European conference on computer vision}, 2020.

\bibitem{MLSYS2022_c48e8203}
\BIBentryALTinterwordspacing
H.~Tang, Z.~Liu, X.~Li, Y.~Lin, and S.~Han, ``Torchsparse: Efficient point cloud inference engine,'' in \emph{Proceedings of Machine Learning and Systems}, D.~Marculescu, Y.~Chi, and C.~Wu, Eds., vol.~4, 2022, pp. 302--315. [Online]. Available: \url{https://proceedings.mlsys.org/paper_files/paper/2022/file/c48e820389ae2420c1ad9d5856e1e41c-Paper.pdf}
\BIBentrySTDinterwordspacing

\bibitem{Tang_2023_CVPR}
H.~Tang, S.~Yang, Z.~Liu, K.~Hong, Z.~Yu, X.~Li, G.~Dai, Y.~Wang, and S.~Han, ``Torchsparse++: Efficient point cloud engine,'' in \emph{Proceedings of the IEEE/CVF Conference on Computer Vision and Pattern Recognition (CVPR) Workshops}, June 2023, pp. 202--209.

\bibitem{8ivfb}
E.~d’Eon, B.~Harrison, T.~Myers, and P.~A. Chou, ``8i voxelized full bodies-a voxelized point cloud dataset,'' \emph{ISO/IEC JTC1/SC29 Joint WG11/WG1 (MPEG/JPEG) input document WG11M40059/WG1M74006}, vol.~7, no.~8, p.~11, 2017.

\bibitem{owlii}
Y.~L. Y.~Xu and Z.~Wen, ``Owlii dynamic human mesh sequence dataset,'' \emph{ISO/IEC JTC1/SC29/WG11 m41658}, 2017.

\bibitem{schnabel2006octree}
R.~Schnabel and R.~Klein, ``Octree-based point-cloud compression.'' \emph{PBG@ SIGGRAPH}, vol.~3, pp. 111--121, 2006.

\bibitem{sze2014high}
V.~Sze, M.~Budagavi, and G.~J. Sullivan, ``High efficiency video coding (hevc),'' in \emph{Integrated circuit and systems, algorithms and architectures}.\hskip 1em plus 0.5em minus 0.4em\relax Springer, 2014, vol.~39, p.~40.

\bibitem{balle2016end}
J.~Ball{\'e}, V.~Laparra, and E.~P. Simoncelli, ``End-to-end optimized image compression,'' \emph{arXiv preprint arXiv:1611.01704}, 2016.

\bibitem{balle2018variational}
J.~Ball{\'e}, D.~Minnen, S.~Singh, S.~J. Hwang, and N.~Johnston, ``Variational image compression with a scale hyperprior,'' \emph{arXiv preprint arXiv:1802.01436}, 2018.

\bibitem{minnenbt18}
D.~Minnen, J.~Ball{\'{e}}, and G.~Toderici, ``Joint autoregressive and hierarchical priors for learned image compression,'' in \emph{Advances in Neural Information Processing Systems 31: Annual Conference on Neural Information Processing Systems 2018, NeurIPS 2018, 3-8 December 2018, Montr{\'{e}}al, Canada}, S.~Bengio, H.~M. Wallach, H.~Larochelle, K.~Grauman, N.~Cesa{-}Bianchi, and R.~Garnett, Eds., 2018, pp. 10\,794--10\,803.

\bibitem{Choy_2019_CVPR}
C.~Choy, J.~Gwak, and S.~Savarese, ``4d spatio-temporal convnets: Minkowski convolutional neural networks,'' in \emph{Proceedings of the IEEE/CVF Conference on Computer Vision and Pattern Recognition (CVPR)}, June 2019.

\bibitem{zhang2022reggeonet}
Q.~Zhang, J.~Hou, Y.~Qian, A.~B. Chan, J.~Zhang, and Y.~He, ``Reggeonet: Learning regular representations for large-scale 3d point clouds,'' \emph{International Journal of Computer Vision}, vol. 130, no.~12, pp. 3100--3122, 2022.

\bibitem{7025414}
C.~Zhang, D.~Florêncio, and C.~Loop, ``Point cloud attribute compression with graph transform,'' in \emph{2014 IEEE International Conference on Image Processing (ICIP)}, 2014, pp. 2066--2070.

\bibitem{7482691}
R.~L. de~Queiroz and P.~A. Chou, ``Compression of 3d point clouds using a region-adaptive hierarchical transform,'' \emph{IEEE Transactions on Image Processing}, vol.~25, no.~8, pp. 3947--3956, 2016.

\bibitem{g_pcc}
{WG 7 and MPEG 3D Graphics Coding}, ``G-pcc codec description v12.'' \emph{ISO/IEC JTC 1/SC 29/WG 7 N00271}, 2022.

\bibitem{9191183}
E.~Pavez, B.~Girault, A.~Ortega, and P.~A. Chou, ``Region adaptive graph fourier transform for 3d point clouds,'' in \emph{2020 IEEE International Conference on Image Processing (ICIP)}, 2020, pp. 2726--2730.

\bibitem{10693649}
Z.~Guo, Y.~Zhang, L.~Zhu, H.~Wang, and G.~Jiang, ``Tsc-pcac: Voxel transformer and sparse convolution-based point cloud attribute compression for 3d broadcasting,'' \emph{IEEE Transactions on Broadcasting}, pp. 1--13, 2024.

\bibitem{jpeg}
WG1., ``Verification model description for jpeg pleno learning-based point cloud coding v4.0,'' \emph{ISO/IEC JTC 1/SC 29/WG1 N100709,}, 2024.

\bibitem{NEURIPS2018_4e8412ad}
\BIBentryALTinterwordspacing
E.~Insafutdinov and A.~Dosovitskiy, ``Unsupervised learning of shape and pose with differentiable point clouds,'' in \emph{Advances in Neural Information Processing Systems}, S.~Bengio, H.~Wallach, H.~Larochelle, K.~Grauman, N.~Cesa-Bianchi, and R.~Garnett, Eds., vol.~31.\hskip 1em plus 0.5em minus 0.4em\relax Curran Associates, Inc., 2018. [Online]. Available: \url{https://proceedings.neurips.cc/paper_files/paper/2018/file/4e8412ad48562e3c9934f45c3e144d48-Paper.pdf}
\BIBentrySTDinterwordspacing

\bibitem{Lin_Kong_Lucey_2018}
\BIBentryALTinterwordspacing
C.-H. Lin, C.~Kong, and S.~Lucey, ``Learning efficient point cloud generation for dense 3d object reconstruction,'' \emph{Proceedings of the AAAI Conference on Artificial Intelligence}, vol.~32, no.~1, Apr. 2018. [Online]. Available: \url{https://ojs.aaai.org/index.php/AAAI/article/view/12278}
\BIBentrySTDinterwordspacing

\bibitem{10.1145/3355089.3356513}
\BIBentryALTinterwordspacing
W.~Yifan, F.~Serena, S.~Wu, C.~\"{O}ztireli, and O.~Sorkine-Hornung, ``Differentiable surface splatting for point-based geometry processing,'' \emph{ACM Trans. Graph.}, vol.~38, no.~6, nov 2019. [Online]. Available: \url{https://doi.org/10.1145/3355089.3356513}
\BIBentrySTDinterwordspacing

\bibitem{Wiles_2020_CVPR}
O.~Wiles, G.~Gkioxari, R.~Szeliski, and J.~Johnson, ``Synsin: End-to-end view synthesis from a single image,'' in \emph{Proceedings of the IEEE/CVF Conference on Computer Vision and Pattern Recognition (CVPR)}, June 2020.

\bibitem{10274683}
M.~You, M.~Guo, X.~Lyu, H.~Liu, and J.~Hou, ``Learning a locally unified 3d point cloud for view synthesis,'' \emph{IEEE Transactions on Image Processing}, vol.~32, pp. 5610--5622, 2023.

\bibitem{Fridovich-Keil_2022_CVPR}
S.~Fridovich-Keil, A.~Yu, M.~Tancik, Q.~Chen, B.~Recht, and A.~Kanazawa, ``Plenoxels: Radiance fields without neural networks,'' in \emph{Proceedings of the IEEE/CVF Conference on Computer Vision and Pattern Recognition (CVPR)}, June 2022, pp. 5501--5510.

\bibitem{Hu_2023_CVPR}
T.~Hu, X.~Xu, R.~Chu, and J.~Jia, ``Trivol: Point cloud rendering via triple volumes,'' in \emph{Proceedings of the IEEE/CVF Conference on Computer Vision and Pattern Recognition (CVPR)}, June 2023, pp. 20\,732--20\,741.

\bibitem{Chang_2023_CVPR}
J.-H.~R. Chang, W.-Y. Chen, A.~Ranjan, K.~M. Yi, and O.~Tuzel, ``Pointersect: Neural rendering with cloud-ray intersection,'' in \emph{Proceedings of the IEEE/CVF Conference on Computer Vision and Pattern Recognition (CVPR)}, June 2023, pp. 8359--8369.

\bibitem{10.1145/3592433}
\BIBentryALTinterwordspacing
B.~Kerbl, G.~Kopanas, T.~Leimkuehler, and G.~Drettakis, ``3d gaussian splatting for real-time radiance field rendering,'' \emph{ACM Trans. Graph.}, vol.~42, no.~4, jul 2023. [Online]. Available: \url{https://doi.org/10.1145/3592433}
\BIBentrySTDinterwordspacing

\bibitem{hearn2004computer}
D.~Hearn, \emph{Computer graphics with OpenGL}.\hskip 1em plus 0.5em minus 0.4em\relax Pearson Education India, 2004.

\bibitem{5980567}
R.~B. Rusu and S.~Cousins, ``3d is here: Point cloud library (pcl),'' in \emph{2011 IEEE International Conference on Robotics and Automation}, 2011, pp. 1--4.

\bibitem{ravi2020pytorch3d}
N.~Ravi, J.~Reizenstein, D.~Novotny, T.~Gordon, W.-Y. Lo, J.~Johnson, and G.~Gkioxari, ``Accelerating 3d deep learning with pytorch3d,'' \emph{arXiv:2007.08501}, 2020.

\bibitem{NEURIPS2019_bdbca288}
\BIBentryALTinterwordspacing
A.~Paszke, S.~Gross, F.~Massa, A.~Lerer, J.~Bradbury, G.~Chanan, T.~Killeen, Z.~Lin, N.~Gimelshein, L.~Antiga, A.~Desmaison, A.~Kopf, E.~Yang, Z.~DeVito, M.~Raison, A.~Tejani, S.~Chilamkurthy, B.~Steiner, L.~Fang, J.~Bai, and S.~Chintala, ``Pytorch: An imperative style, high-performance deep learning library,'' in \emph{Advances in Neural Information Processing Systems}, H.~Wallach, H.~Larochelle, A.~Beygelzimer, F.~d\textquotesingle Alch\'{e}-Buc, E.~Fox, and R.~Garnett, Eds., vol.~32.\hskip 1em plus 0.5em minus 0.4em\relax Curran Associates, Inc., 2019. [Online]. Available: \url{https://proceedings.neurips.cc/paper_files/paper/2019/file/bdbca288fee7f92f2bfa9f7012727740-Paper.pdf}
\BIBentrySTDinterwordspacing

\bibitem{10.1145/1103900.1103907}
\BIBentryALTinterwordspacing
M.~Alexa, M.~Gross, M.~Pauly, H.~Pfister, M.~Stamminger, and M.~Zwicker, ``Point-based computer graphics,'' in \emph{ACM SIGGRAPH 2004 Course Notes}, ser. SIGGRAPH '04.\hskip 1em plus 0.5em minus 0.4em\relax New York, NY, USA: Association for Computing Machinery, 2004, p. 7–es. [Online]. Available: \url{https://doi.org/10.1145/1103900.1103907}
\BIBentrySTDinterwordspacing

\bibitem{rosenthal2008image}
P.~Rosenthal and L.~Linsen, ``Image-space point cloud rendering,'' in \emph{Proceedings of Computer Graphics International}, 2008, pp. 136--143.

\bibitem{SAINZ2004869}
\BIBentryALTinterwordspacing
M.~Sainz and R.~Pajarola, ``Point-based rendering techniques,'' \emph{Computers \& Graphics}, vol.~28, no.~6, pp. 869--879, 2004. [Online]. Available: \url{https://www.sciencedirect.com/science/article/pii/S0097849304001530}
\BIBentrySTDinterwordspacing

\bibitem{7413904}
M.~Schütz and M.~Wimmer, ``High-quality point-based rendering using fast single-pass interpolation,'' in \emph{2015 Digital Heritage}, vol.~1, 2015, pp. 369--372.

\bibitem{vae}
D.~P. Kingma and M.~Welling, ``Auto-encoding variational bayes,'' \emph{arXiv preprint arXiv:1312.6114}, 2013.

\bibitem{MAGGIORDOMO2020101943}
\BIBentryALTinterwordspacing
A.~Maggiordomo, F.~Ponchio, P.~Cignoni, and M.~Tarini, ``Real-world textured things: A repository of textured models generated with modern photo-reconstruction tools,'' \emph{Computer Aided Geometric Design}, vol.~83, p. 101943, 2020. [Online]. Available: \url{https://www.sciencedirect.com/science/article/pii/S0167839620301308}
\BIBentrySTDinterwordspacing

\bibitem{dai2017scannet}
A.~Dai, A.~X. Chang, M.~Savva, M.~Halber, T.~Funkhouser, and M.~Nie{\ss}ner, ``Scannet: Richly-annotated 3d reconstructions of indoor scenes,'' in \emph{Proc. Computer Vision and Pattern Recognition (CVPR), IEEE}, 2017.

\bibitem{bdrate}
G.~Bjøntegaard, ``Calculation of average psnr differences between rdcurves,'' \emph{ITU-T SG 16/Q6, 13th VCEG Meeting. document VCEGM33}, April 2001.

\bibitem{yuv611}
L.~Cruz, ``Jpeg pleno point cloud coding common training and test conditions v1.3,'' \emph{ITU-T SG 16/Q6, ISO/IEC JTC 1/SC 29/WG1, N100276}, July 2022.

\bibitem{zou2022devil}
R.~Zou, C.~Song, and Z.~Zhang, ``The devil is in the details: Window-based attention for image compression,'' in \emph{Proceedings of the IEEE/CVF conference on computer vision and pattern recognition}, 2022, pp. 17\,492--17\,501.

\bibitem{li2021deep}
J.~Li, B.~Li, and Y.~Lu, ``Deep contextual video compression,'' \emph{Advances in Neural Information Processing Systems}, vol.~34, pp. 18\,114--18\,125, 2021.

\bibitem{sheng2022temporal}
X.~Sheng, J.~Li, B.~Li, L.~Li, D.~Liu, and Y.~Lu, ``Temporal context mining for learned video compression,'' \emph{IEEE Transactions on Multimedia}, vol.~25, pp. 7311--7322, 2022.

\bibitem{cheng2020learned}
Z.~Cheng, H.~Sun, M.~Takeuchi, and J.~Katto, ``Learned image compression with discretized gaussian mixture likelihoods and attention modules,'' in \emph{Proceedings of the IEEE/CVF conference on computer vision and pattern recognition}, 2020, pp. 7939--7948.

\end{thebibliography}



\begin{IEEEbiography}
[{\includegraphics[width=1in,height=1.2in,clip,keepaspectratio]{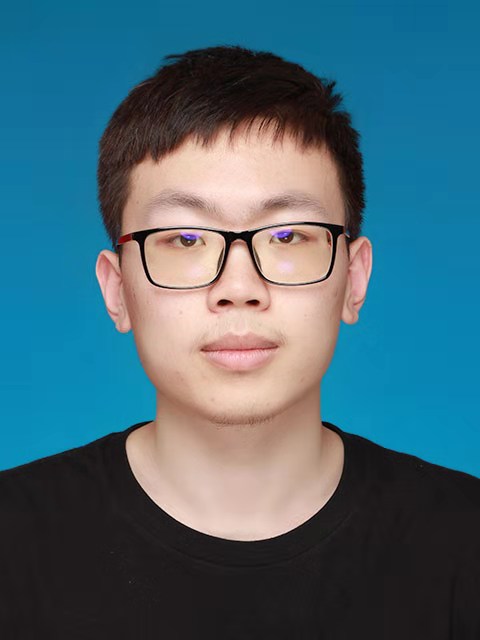}}]{Xiao Huo} received the B.E. degree from Xidian University, Xi’an, China, in 2019, where he is currently pursuing the Ph.D. degree in Communication Engineering. His current research interests include point cloud compression, light field camera calibration, light
field imaging, and 3-D reconstruction.
\end{IEEEbiography}

\begin{IEEEbiography}
[{\includegraphics[width=0.95in,height=1.15in,clip,keepaspectratio]{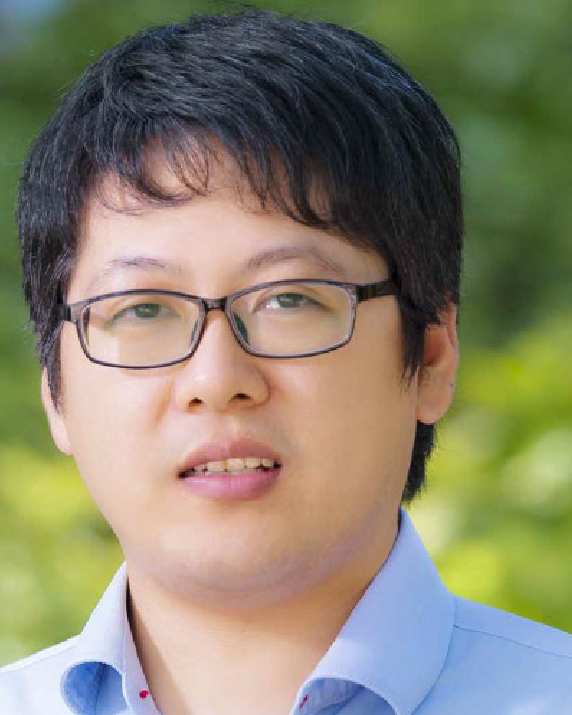}}]{Junhui Hou}(Senior Member, IEEE) is an Asso-
ciate Professor with the Department of Computer
Science, City University of Hong Kong. He holds a
B.Eng. degree in information engineering (Talented
Students Program) from the South China Univer-
sity of Technology, Guangzhou, China (2009), an
M.Eng. degree in signal and information processing
from Northwestern Polytechnical University, Xi’an,
China (2012), and a Ph.D. degree from the School
of Electrical and Electronic Engineering, Nanyang
Technological University, Singapore (2016). His re-
search interests are multi-dimensional visual computing.

Dr. Hou received the Early Career Award (3/381) from the Hong Kong
Research Grants Council in 2018 and the NSFC Excellent Young Scientists
Fund in 2024. He has served or is serving as an Associate Editor for IEEE
Transactions on Visualization and Computer Graphics, IEEE Transactions on
Image Processing, IEEE Transactions on Multimedia, and IEEE Transactions
on Circuits and Systems for Video Technology.
\end{IEEEbiography}

\begin{IEEEbiography}
[{\includegraphics[width=1in,height=1.2in,clip,keepaspectratio]{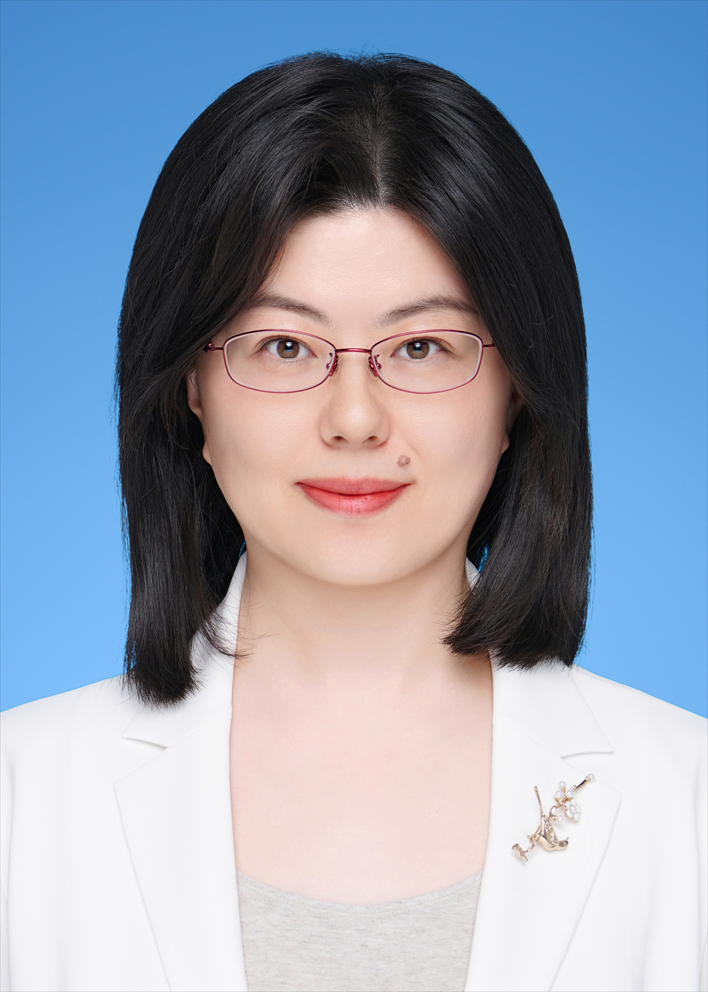}}]{Shuai Wan} (Member, IEEE) received the B.E. degree in Telecommunication Engineering and the M.E. degree in Communication and Information System from Xidian University, Xi’an, China, in 2001 and 2004, respectively, and obtained the Ph.D. in Electronic Engineering from Queen Mary, University of London in 2007. She is now a Professor in Northwestern Polytechnical University, Xi’an, China. Her research interests include scalable/multiview video coding, video quality assessment and hyperspectral image compression.
\end{IEEEbiography}

\begin{IEEEbiography}
[{\includegraphics[width=1in,height=1.2in,clip,keepaspectratio]{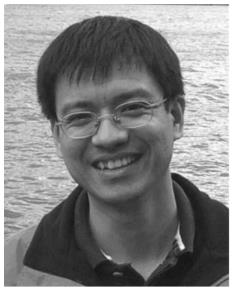}}]{Fuzheng Yang}(Member, IEEE) received the B.E. degree in Telecommunication Engineering, the M.E. degree and the Ph.D. in Communication and Information System from Xidian University, Xi’an, China, in 2000, 2003 and 2005, respectively. He became a lecturer and an Associate Professor in Xidian University in 2005 and 2006, respectively. He has been a professor of communications engineering with Xidian University since 2012. He is also an Adjunct Professor of School of Engineering in RMIT University. During 2006-2007, he served as a visiting scholar and postdoctoral researcher in Department of Electronic Engineering in Queen Mary, University of London. His research interests include video quality assessment, video coding and multimedia communication.
\end{IEEEbiography}

\vspace{11pt}


\vfill

\end{document}